\documentstyle[12pt]{article}

\newcommand{\tr}{{\rm tr}}
\newcommand{\Tr}{{\rm Tr}}
\newcommand{\Det}{{\rm Det}}
\title{
First and Second Order Phase Transitions
\\ in Maxwell--Chern-Simons Theory
\\ Coupled to Fermions
\thanks{To be published in Intern. J. Mod. Phys. A.}
}
\author{%
  Kei-Ichi Kondo
  \thanks{e-mail: kondo@cuphd.nd.chiba-u.ac.jp}
  \vspace{1.2em}
  \\
  Department of Physics, Faculty of Science \\
         Chiba University, Chiba 263, Japan
  \thanks{Permanent  address.
  Address from September 1995  to February 1996:
  HLRZ c/o KFA J\"ulich, D-52425 J\"ulich, Germany,
  and
  address from March 1996  to December 1996:
  Department of Physics, Theoretical Physics,
  University of Oxford,
  1 Keble Road, Oxford, OX1 3NP, UK.}
}
\date{CHIBA-EP-89, \\
April 1995, \\
hep-ph/9509345}
\begin{document}
\maketitle
\centerline{Abstract}
In the Maxwell--Chern-Simons theory coupled to $N_f$ flavors
of 4-component fermions (or even number of 2-component
fermions) we construct the gauge-covariant effective
potential written in terms of two order parameters which are
able to probe the breakdown of chiral symmetry and parity.
In the absence of the bare Chern-Simons term, we show that
the chiral symmetry is spontaneously broken for fermion
flavors $N_f$ below a certain finite critical number
$N_f^c$, while the  parity is not broken spontaneously.
This chiral phase transition is of the second order.
In the presence of the bare Chern-Simons term, on  the
other hand, the chiral phase transition associated  with the
spontaneous breaking  of chiral symmetry is shown to
continue to exist, although the parity is explicitly broken.
However it is shown that the existence of the bare
Chern-Simons term changes the order of the chiral transition
into the  first order, no matter how small the bare
Chern-Simons coefficient  may be.  This gauge-invariant
result is consistent with that recently obtained by the
Schwinger-Dyson equation  in the non-local gauge.

\newpage
\section{Introduction}
\setcounter{equation}{0}
The Chern-Simons term \cite{JT81,Schonfeld81,DJT81}
generates a gauge-invariant mass for the gauge field,
combined with the ordinary Maxwell term. Formally, in the
Chern-Simons theory which does not include the kinetic term
for the gauge field, the gauge sector of the theory is
strictly renormalizable, although the Maxwell theory is
superrenormalizable in 2+1 space-time dimensions.
The  Chern-Simons term affects the long-distance behavior
of the theory and the perturbative expansion of
Chern-Simons theory contains logarithmic divergences
\cite{Wilczek90,Fradkin91,DM92}.
However, the topological nature of the theory should allow
only trivial, finite renormalization \cite{Wilczek90}.
\par
Actually it is verified that this happens in the various
renormalization scheme.
The massive matter fields may have a finite one-loop
radiative correction to the statistical parameter $\theta$,
the coefficient of the Chern-Simons term.
It is generally known as the Coleman-Hill
nonrenormalization theorem
\cite{CH85} that the topological mass term at one-loop
order is the only one radiative correction and there are no
radiative corrections from higher-order loops to the
topological mass term. Therefore the coefficient of the
topological mass term at one-loop order is exact to all
higher orders.  The requirement of this nonrenormalization
theorem is the gauge invariance and analyticity of the
one-loop $n$-point functions.
\par
Indeed, from the explicit consideration of the Feynman
graph, the following results are obtained. When there is no
topological mass at the tree level, no topological mass
terms are induced radiatively to the two-loop order in
either Abelian or non-Abelian gauge theories with {\it
massive} fermions in 2+1 space-time dimensions
\cite{KS85}.
Moreover, if there is a nonzero tree-level topological
mass, the radiative correction to the topological mass is
shown to vanish in Abelian gauge theory coupled to the
massive fermion \cite{BL85}.
However it should be remarked that the existence of
{\it massless} charged particles or spontaneous
gauge symmetry breaking violates the initial assumptions of
the Coleman-Hill theorem and higher loop renormalizations
of the Chern-Simons term appear
\cite{SSW89,ST91,Chen90}.
\par
The nonrenormalization theorem in the weak coupling
expansion can be extended to the
large flavor ($N_f$) case, i.e., to the $1/N_f$ expansion:
there is no  radiative corrections to the Chern-Simons
coefficient
$\theta$ in the $1/N$ expansion \cite{CHLP93}.
This argument do not apply to the case in which the
fermions remain massless due to absence of the
analyticity.
 In the Abelian case,
the nonrenormalization theorem to all orders is also shown
by using the effective potential \cite{Lee86}.
It will be interesting to extend these analyses to the
non-Abelian case \cite{CSW92} such as QCD3
\cite{AP81,AN90}.
\par
In this paper we adopt the $1/N_f$ expansion as a
systematic truncation scheme.  The $1/N_f$ expansion is
quite interesting from the viewpoint of the
non-perturbative study, since a class of theories with a
four-fermion interaction is shown to be renormalizable in
the framework of $1/N_f$ expansion
\cite{RWP89} in spite of its nonrenormalizability in the
weak coupling expansion.
The Chern-Simons term may affect the high-energy behavior
of the gauged four-fermion model \cite{CHLP93,Chen94}
and the pattern of dynamical generation for the fermion
mass, although the four-fermion interaction is
not investigated in this paper.
\par
In this paper we regard the dimensional coupling constant
$e$, the electric charge, as a genuine dimensional
parameter that sets the natural scale of the theory.
This viewpoint has been taken by Appelquist et al.
\cite{ABKW86} and several authors.   The dynamically
generated fermion will have a mass $m_d$ which is
proportional to this scale: $m_d
\sim e^2$. The fermion mass is generated when $1/N_f >
1/N_f^c$ where $N_f^c$ is called the critical number of
fermion flavors.  Another way of viewing the dimensional
parameter is to take it as the ultraviolet (UV) cutoff of
the theory where $\Lambda$ is the UV cutoff and
$g=e^2/\Lambda$ is a dimensionless coupling \cite{GN74}.  In
this case, the only dimensional parameter in the model is
the UV cutoff.  In this standpoint, if the dynamical mass is
generated, it will be independent of the UV cutoff
$\Lambda$.   This leads to the non-perturbative
renormalization for
$g$, i.e., the coupling is running, i.e.,
$\beta(g) := \Lambda {dg \over d\Lambda} \not= 0$.
Therefore, the first viewpoint is
not compatible with the second one and two viewpoints are
in many ways different from each other.
\par
\par
The Maxwell-Chern-Simons theory
\cite{JT81,Schonfeld81,DJT81,Wilczek90,Fradkin91,DM92}
with $N_f$ flavors of Dirac fermions in (2+1)-dimensions is
defined by the following euclidean Lagrangian density:
\begin{eqnarray}
{\cal L}  &=&
  {\beta \over 4} F_{\mu\nu}^2
+ {i \theta \over 2} \epsilon_{\mu \nu \rho} A_\mu
\partial_\nu A_\rho + {\cal L}_{GF}
\nonumber\\&&
+ \bar \Psi^\alpha (i \gamma_\mu \partial_\mu - m_e - m_o
\tau) \Psi^\alpha
+ e \bar \Psi^\alpha \gamma_\mu \Psi^\alpha A_\mu ,
\label{lagrangian}
\end{eqnarray}
where the index $\alpha$ denotes the fermion flavors
($\alpha = 1, ..., N_f)$
and
${\cal L}_{GF}$ is the gauge-fixing term specified in the
next section.
In the above Lagrangian density we have
introduced an interpolating parameter
$\beta$ between the Maxwell $(\beta = 1, \theta = 0)$ and
the Chern-Simons $(\beta = 0, \theta\not= 0)$ theories.
Our analysis based on the effective potential in this
paper can be applied equally to the case of $\beta=0$ as
well as $\beta\not=0$.
The (2+1)-dimensional QED (QED3) with a Chern-Simons term is
a special case of this model.
For the Chern-Simons theory, the term
${\beta \over 4} F_{\mu\nu}^2$
can be interpreted as a regularization which is removed
finally, i.e. $\beta \rightarrow 0$ in this case.
\par
 We start from the 4-component formulation for fermions in
(2+1)-dimensions \cite{ABKW86}.
Each 4-component fermion
\footnote{
We use $\Psi$ without the index $\alpha$ to denote each
4-component fermion.  }
$\Psi$
is decomposed into two 2-component fermions
$\psi_1, \psi_2$ as
\begin{equation}
 \Psi =   {\psi_1  \choose \psi_2 } .
\end{equation}
Then the Dirac conjugate
$
\bar \Psi := \Psi^\dagger \gamma_0
$
is written as
\footnote{In what follows we use the notation $A:=B$ in the
sense that $A$ is defined by $B$, while $A \equiv B$ implies
that $A$ is identically equal to $B$.}
\begin{equation}
\bar \Psi  = (\bar \psi_1, - \bar \psi_2)
\end{equation}
by using
$\bar \psi_a := \psi_a^\dagger \sigma_3 (a=1,2)$, see
Appendix A.
Accordingly the gamma matrices satisfying the Clifford
algebra
$\{\gamma_\mu, \gamma_\nu\} = -2 \delta_{\mu\nu}$ are
written by 4 $\times$ 4 matrices.
In this reducible representation,  there exist 4 $\times$ 4
matrices anticommuting with all the gamma matrices, say
$\gamma_3, \gamma_5$, in  sharp contrast to the
irreducible representation corresponding to
2-component fermions \cite{ABKW86}.   Then we can define the
chiral symmetry similarly as in the (3+1)-dimensional case:
\begin{equation}
\Psi \rightarrow \exp (i\vartheta \gamma_3) \Psi ,
\quad
\bar \Psi \rightarrow \bar \Psi \exp (i\vartheta \gamma_3),
\end{equation}
under which $\bar \Psi i \gamma_\mu \partial_\mu \Psi$ is
invariant.
Moreover, we can define the parity transformation:
\begin{eqnarray}
 \Psi(x_0,\vec x) &\rightarrow&
 {\cal P} \Psi(x_0,\vec x_P) ,
 \nonumber\\
 \bar \Psi(x_0,\vec x) &\rightarrow&
 \bar \Psi(x_0,\vec x_P){\cal P} ,
 \nonumber\\
 A_\mu(x_0,\vec x) &\rightarrow&
 (-1)^{\delta_{\mu1}} A_\mu(x_0,\vec x_P) ,
 \label{parity}
\end{eqnarray}
where $\vec x_P=(-x_1,x_2)$ for the spacial coordinate
$\vec x=(x_1,x_2)$ and
${\cal P}=-i\gamma_5 \gamma_1$.
The parity transformation~(\ref{parity}) is equivalent to
$
\psi_1(x_0,\vec x) \rightarrow
\sigma_1 \psi_2(x_0,\vec x_P)
$
and
$
\psi_2(x_0,\vec x) \rightarrow
\sigma_1 \psi_1(x_0,\vec x_P) ,
$
for 2-component fermions.
The transformation property of each term in the
Lagrangian under two transformations is summarized as
follows.
\par
\begin{center}
 \begin{tabular}{|l||c|c|}
 \hline
   \multicolumn{3}{|c|}{Transformation property}   \\
  \hline
    term                 &  chiral & parity \\
   \hline \hline
   $\bar \Psi \Psi$      &     odd & even \\
   \hline
   $\bar \Psi \tau \Psi$ &    even & odd \\
   \hline
   $\epsilon_{\mu \nu \rho} A_\mu \partial_\nu A_\rho$ &
   even & odd \\
   \hline
 \end{tabular}
\end{center}

\par
The  order parameter for chiral symmetry breaking is
defined by
\begin{equation}
   \langle \bar \Psi  \Psi \rangle
 \equiv \langle \bar \psi_1  \psi_1 \rangle
   - \langle \bar \psi_2  \psi_2 \rangle,
\end{equation}
while the order parameter for parity breaking
is defined by
\begin{equation}
   \langle \bar \Psi  \tau \Psi \rangle
 \equiv \langle \bar \psi_1  \psi_1 \rangle
   + \langle \bar \psi_2  \psi_2 \rangle.
\end{equation}
\par
According to analytical and
numerical studies \cite{KM95} of the Schwinger-Dyson equation
for QED3 with a Chern-Simons term
$(\beta = 1, \theta \not= 0, m_e = m_o = 0)$,
we can write the schematic phase diagram for the
Maxwell-Chern-Simons theory with $N_f$ flavors of
4-component fermions, see Fig.~1.  First note that, when
$\theta\not=0$, the parity is explicitly broken,
$
 \langle \bar \Psi \tau \Psi \rangle \not= 0.
$
The solid line in Fig.~1 denotes the critical line of
chiral phase transition below which the chiral symmetry is
spontaneously broken,
$
 \langle \bar \Psi \Psi \rangle \not= 0.
$
In Fig.~1 we have chosen typical five points, assuming
that the critical number of flavors $N_f^c$ is nonzero and
finite:
$0 < N_f^c < \infty$. The points A and E in Fig.~1 are in the
chiral symmetric phase; the point A is in the parity
even phase, while E is in the parity odd phase. The point D
is located just on the critical line.
The points B and C are in the chiral-symmetry-breaking
phase; the point B is in the parity even phase, while C
is in the parity odd phase.
For the order of phase transition, it is well known that the
point $(N_f,\theta) = (N_f^c,0)$ exhibits the continuous
second order transition through the
analyses of the Schwinger-Dyson equation
\cite{ANW88}.
\par

\par
By using the Schwinger-Dyson equation,
the critical number of flavors $N_f^c$ has been evaluated.
Finite and nonzero critical number, $0<N_f^c<\infty$ has been
reported:
$N_f^c={32 \over \pi^2} \cong 3.2$ first in the Landau
gauge by  Appelquist, Nash and Wijewardhana \cite{ANW88},
and $N_f^c={128 \over 3\pi^2} \cong 4.3$ in the
heuristic treatment of the non-local gauge by Nakatani
\cite{Nakatani88} and $N_f^c \cong 3 \sim 4$ by Nash
\cite{Nash89}.  These results are confirmed by
the systematic investigation of the non-local gauge by
Kondo et al. \cite{KEIT95} and by Kondo and Maris
\cite{KM95}.
On the other hand, infinity of $N_f$,
$N_f=\infty$, was claimed by
\cite{PW88,AJP88} and subsequent works
\cite{CP91,CPW92}.
Different approximations to the vertex function and the
existence of infrared cutoff may lead to different results,
see \cite{AJM90} and \cite{KN92}. Such a controversy exists
\cite{MMV88,Matsuki91} also in the approach of the
Cornwall-Jackiw-Tomboulis (CJT) effective potential
\cite{CJT74}.
The Monte Carlo simulation of non-compact lattice QED3
shows $N_f^c=3.5 \pm 0.5$ according to
Dagotto, Kogut and Kocic \cite{DKK89}.
The result are controversial up to now.
This shows that it is extremely difficult to find a
consistent truncation scheme in the Schwinger-Dyson
approach which is able to deduce the gauge-invariant result.
For more details on the Schwinger-Dyson equation of QED3,
see for example a review \cite{RW94}.

\par
In the absence of the bare Chern-Simons term, the fermion
mass is dynamically generated for
$N_f<N_f^c$ in such a way that the parity
is preserved, i.e.,
$
 \langle \bar \Psi \Psi \rangle \not= 0
$
and
$
 \langle \bar \Psi \tau \Psi \rangle = 0.
$
This is rephrased  in terms of the fermion dynamical mass
$m_1, m_2$ for two 2-component fermions, $\psi_1, \psi_2$,
respectively. Here the chiral symmetry breaking,
$
 \langle \bar \Psi \Psi \rangle \not= 0 ,
$
implies $m_1\not=m_2$ and no parity  breaking,
$
 \langle \bar \Psi \tau \Psi \rangle = 0,
$
implies more specific pattern for the fermion dynamical
mass:
$$
 \underbrace{m,m,...,m}_{N_f},
 \underbrace{-m,-m,...,-m}_{N_f}.
$$
This result is consistent with various analyses so far
\cite{ABKW86b,Polychronakos88,RY86,CCW91,SW92,KEIT95}
as well as the general consideration by Vafa and Witten
\cite{VW84}.

\par
A new and remarkable feature discovered in \cite{KM95} is
that the order of chiral phase transition  turns into
the first order when
$\theta\not=0$.  Namely the chiral phase transition is of
the first order on the whole critical line except for one
point $(N_f,\theta) = (N_f^c,0)$ in the phase diagram, which
is in sharp contrast to the previous analysis
\cite{HP93}.
The meaning of the first order transition
is explained from the behavior of dynamically generated
fermion mass as shown in Fig.~2.  In Fig.~2, fermion
masses $m_1, m_2$ are shown as different functions of
$\theta$ when $N_f<N_f^c$. At the point B,
$\theta=0$ and fermion masses for two
2-component fermions are generated in such a way that  the
parity is preserved:
$m_1=-m_2>0$
(we choose a convention such that $m_1>0$ without loss
of generality) and the chiral symmetry is spontaneously
broken,
$
 \langle \bar \Psi \Psi \rangle \not= 0.
$
When $\theta\not=0$, we see this pattern for mass
generation is destroyed so that
$|m_1|\not=|m_2|$ ($m_1>0, m_2<0$). As $\theta$ is
increased while $N_f$ is kept fixed, each $m_a(a=1,2)$
increases monotonically.  However, at
$\theta=\theta_c$,
$m_2(\theta)$ jumps discontinuously towards $m_1(\theta)$.
For $\theta>\theta_c$, it is realized that
$m_1(\theta)=m_2(\theta)$, which implies restoration of the
chiral symmetry:
$
 \langle \bar \Psi \Psi \rangle = 0.
$
In this sense the chiral phase transition is of the first
order when $\theta\not=0$.

\par
However this result is available only in a specific gauge.
In the Schwinger-Dyson equation approach \cite{KM95}, the
non-local gauge was adopted in order to avoid the vertex
correction.  A possible way to improve the gauge-dependence
in the Schwinger-Dyson equation approach is to improve the
vertex so as to be consistent with the Ward-Takahashi
identity.  However this way is not unique.

\par
The purpose of this paper is show the existence of the first
order phase transition in the presence of the
bare Chern-Simons term as well as the second order
transition in the absence of the bare Chern-Simons term
in the gauge-invariant way.  For this, we construct the
{\it gauge-covariant} (gauge-parameter independent)
effective potential in terms of two order parameters:
$
 \phi = \langle \bar \Psi \Psi \rangle
$
and
$
 \chi = \langle \bar \Psi \tau  \Psi \rangle,
$
up to the leading order of $1/N_f$ expansion.
To construct the effective potential, we adopt the
inversion method
\cite{Fukuda88,UKF90}.

\par
This paper is organized as follows.
In section 2, we show how the gauge-covariant result is
obtained for the gauge-invariant order parameters through
the evaluation of the effective potential in the scheme of
$1/N_f$ expansion. In section 3, after introducing two
external sources which trigger the spontaneous breaking of
the chiral symmetry and parity, we show how explicitly the
order parameter is related to  the vacuum polarization of the
photon propagator.
In section 4, we observe the dependence of the order
parameter on two external sources in detail.
Sections 2 to 4 are preparations for performing the
inversion.
Now, in section 5, we state a general strategy of obtaining
the non-trivial solutions through the inversion method in
the presence of two external sources.  In section 6, as a
special case we study the system with a single order
parameter and in a slightly simplified situation we show
that the chiral symmetry is spontaneously broken in the
three-dimensional gauge theory, e.g., QED3 by using the
inversion method. In section 7, we give the argument on the
behavior of the effective potential near the stationary
point in the absence of the Chern-Simons term. The
stationary point corresponds to the non-trivial solution
(order parameter) corresponding to the spontaneous breakdown
of the relevant symmetry. In section 8, we study the effect
of the Chern-Simons term on the chiral phase transition by
extending the effective potential given in section 7.
The final section is devoted to conclusion and discussion.
In Appendix A, we give a formulation of 4-component fermion
in (2+1)-dimensions. In Appendix B, we derive the gauge
covariant formula in the $1/N_f$ expansion given in section 2
for the gauge-invariant order parameter. In Appendix C, the
vacuum polarization tensor in three-dimensional  gauge
theory is obtained  together with its power-series expansion
in the external source.

\section{Order parameter in the $1/N_f$ expansion}
\setcounter{equation}{0}
In this paper we
take the covariant gauge-fixing:
\begin{eqnarray}
{\cal L}_{GF}  =  {1 \over 2\xi} (\partial_\mu A_\mu)^2 .
\label{GF}
\end{eqnarray}
{}From the Lagrangian~(\ref{lagrangian}) with the
gauge-fixing term~(\ref{GF}), it is easy to see that the
inverse  of the bare photon propagator reads
\begin{eqnarray}
 D_{\mu \nu}^{(0)}{}^{-1}(k)
 =  \beta k^2 \left( \delta_{\mu \nu}  - {k_\mu k_\nu \over
k^2}  \right)
   +  {1 \over \xi} k_\mu k_\nu
 +  \theta \epsilon_{\mu \nu \rho} k_\rho ,
\end{eqnarray}
and hence the bare photon propagator is given by
\begin{eqnarray}
 D_{\mu \nu}^{(0)}(k)   =   {\beta \over \beta^2
k^2+\theta^2}
\left( \delta_{\mu \nu} - {k_\mu k_\nu \over k^2} \right)
   + {k_\mu k_\nu \over k^2} {\xi \over k^2}
 -   {\epsilon_{\mu \nu \rho} k_\rho \over k^2}
 {\theta \over \beta^2 k^2+\theta^2},
\end{eqnarray}
where we have used $\epsilon_{012}=1$ and
$\epsilon_{\mu \nu \lambda} \epsilon_{\rho \sigma \lambda}
= \delta_{\mu \rho} \delta_{\nu \sigma} -\delta_{\mu
\sigma} \delta_{\nu \rho} $.
\par
The general form of the vacuum polarization tensor in
three-dimensional gauge theory is written  as
\begin{eqnarray}
\Pi_{\mu \nu}(k) =   \left[
\delta_{\mu \nu}-{k_\mu k_\nu \over k^2} \right] \Pi_T(k)
+ \epsilon_{\mu\nu\rho} {k_\rho \over \sqrt{k^2}} \Pi_O(k).
\end{eqnarray}

\par
The full photon propagator $D_{\mu \nu}(k)$ is related to
the vacuum polarization tensor $\Pi_{\mu \nu}(k)$ and the
bare photon propagator $D_{\mu \nu}^{(0)}(k)$ through the
Schwinger-Dyson equation:
\begin{eqnarray}
 D_{\mu \nu}{}^{-1}(k)
 =  D_{\mu \nu}^{(0)}{}^{-1}(k) - \Pi_{\mu \nu}(k) .
 \label{SD}
\end{eqnarray}
Therefore the full photon propagator is written in the
form
\begin{eqnarray}
D_{\mu \nu}(k) =   \left[
\delta_{\mu \nu}-{k_\mu k_\nu \over k^2} \right] D_T(k)  +
{k_\mu k_\nu \over k^2} D_L(k) + \epsilon_{\mu \nu \rho}
{k_\rho \over \sqrt{k^2}} D_O(k),
\end{eqnarray}
with
\begin{eqnarray}
 D_T(k)   &=&  {\beta k^2 - \Pi_T(k) \over
 [\beta k^2-\Pi_T(k)]^2+[\theta \sqrt{k^2}-\Pi_O(k)]^2},
 \nonumber\\
 D_O(k)   &=&  {\theta \sqrt{k^2}-\Pi_O(k)
\over
 [\beta k^2-\Pi_T(k)]^2+[\theta \sqrt{k^2}-\Pi_O(k)]^2},
 \nonumber\\
 D_L(k)   &=&   {\xi \over k^2} .
\end{eqnarray}

\par
In order to calculate the order parameter, we introduce the
source term
\begin{eqnarray}
 {\cal L}_J = J_e \bar \Psi^\alpha(x) \Psi^\alpha(x)
 + J_o \bar \Psi^\alpha(x) \tau \Psi^\alpha(x).
 \label{source}
\end{eqnarray}
Then the partition function in the presence of the
source is defined by
\begin{eqnarray}
 Z[J_e, J_o] = \int
 {\cal D}\bar \Psi {\cal D}\Psi {\cal D}A_\mu \exp \left[
 - \int d^3x ({\cal L}+{\cal L}_J) \right],
\end{eqnarray}
and the generating functional is given by
\begin{eqnarray}
 W[J_e, J_o] = - \ln Z[J_e, J_o].
\end{eqnarray}
The order parameter is obtained by
differentiating the generating functional with respect
to the corresponding source.  For the
translation-invariant order parameter, we obtain
\begin{eqnarray}
 \langle \bar \Psi^\alpha \Psi^\alpha \rangle
 &=& {1 \over \Omega}
 {\delta \over \delta J_e} W[J_e, J_o] \Big|_{J_e=J_o=0} ,
\nonumber\\
 \langle \bar \Psi^\alpha \tau \Psi^\alpha \rangle
 &=&  {1 \over \Omega}
 {\delta \over \delta J_o} W[J_e, J_o]\Big|_{J_e=J_o=0},
\end{eqnarray}
where $\Omega = \int d^3x$ is the space-time volume.
The generating functional is given by the sum of
all the bubble diagrams which is depicted in Fig.~3 up
to the leading order of $1/N_f$ expansion.
Separating the bare (interaction free) part with the
superscript
$(0)$, the order parameter which we calculate can be written
in the form (see Appendix B):
\begin{eqnarray}
\langle \bar \Psi \Psi \rangle
&:=& {\langle \bar \Psi^\alpha \Psi^\alpha \rangle \over
N_f}
= \langle \bar \Psi \Psi \rangle^{(0)}
 + {1 \over 2}  N_f^{-1}
\int {d^3k \over (2\pi)^3} D_{\mu \nu}^{(1)}(k)
{\partial \over \partial J_e} \Pi_{\mu \nu}^{(1)}(k),
\nonumber\\
\langle \bar \Psi \tau \Psi \rangle
&:=& {\langle \bar \Psi^\alpha \tau \Psi^\alpha \rangle
\over N_f}
= \langle \bar \Psi \tau \Psi \rangle^{(0)}
 + {1 \over 2}  N_f^{-1}
\int {d^3k \over (2\pi)^3} D_{\mu \nu}^{(1)}(k)
{\partial \over \partial J_o} \Pi_{\mu \nu}^{(1)}(k).
\label{formula}
\end{eqnarray}
Here $\Pi_{\mu \nu}^{(m)}$ denotes the vacuum
polarization in which all the radiative corrections up
to $m$-(fermion) loops are included, and
$D_{\mu \nu}^{(n)}$ denotes the photon propagator
which is constructed from Eq.(\ref{SD}) by using  $\Pi_{\mu
\nu}^{(n)}$.
We use the convention:
$\Pi_T^{(0)}(k) \equiv 0 \equiv \Pi_O^{(0)}(k)$
and  $D_{\mu\nu}^{(0)}$ is the bare photon propagator.
\par
To carry out the calculation in the $1/N_f$ scheme,
therefore, we need to calculate
\begin{eqnarray}
&&  {1 \over 2} \int {d^3k \over (2\pi)^3}
 D_{\mu \nu}^{(n)}(k)
{\partial \over \partial J} \Pi_{\mu \nu}^{(m)}(k)
\nonumber\\
&=& \int {d^3k \over (2\pi)^3} \left[ D_T^{(n)}(k)
{\partial \over \partial J} \Pi_T^{(m)}(k)
+ D_O^{(n)}(k)
{\partial \over \partial J} \Pi_O^{(m)}(k) \right]
\nonumber\\  &=&
\int {d^3k \over (2\pi)^3}
{[\beta k^2-\Pi_T^{(n)}(k)]
{\partial \over \partial J} \Pi_T^{(m)}(k)
+ [\theta \sqrt{k^2}-\Pi_O^{(n)}(k)]
{\partial \over \partial J} \Pi_O^{(m)}(k)
\over   [\beta k^2-\Pi_T^{(n)}(k)]^2
+[\theta \sqrt{k^2}-\Pi_O^{(n)}(k)]^2},
\label{expression}
\end{eqnarray}
where we have used
$\epsilon_{\mu\nu\alpha} \epsilon_{\mu\nu\beta}
= 2\delta_{\alpha \beta}$.
\par
Up to the leading order of $1/N_f$ expansion, we have
only  to consider the case: $n=1, m=1$, which is shown in
Appendix B.  This is a novel observation in this paper. This
should be compared with the usual perturbation theory with
respect to the coupling constant $e^2$.  In this case, the
order of the graph is counted with respect to the coupling,
$e^2$.
In the lowest order of the perturbation series with respect
to the coupling constant, we need to choose
$n=0, m=1$ as is well known in the four-dimensional case
$D=4$ \cite{UKF90}.
It should be remarked that the expression~(\ref{expression})
is free from the gauge-fixing parameter $\xi$.  Therefore,
if we calculate the gauge-invariant order parameter
according to this expression, we will get the
gauge-covariant (i.e. gauge-parameter independent) result.

\par
This should be compared with the Schwinger-Dyson equation
approach in which a specific gauge has to be chosen even in
calculating the gauge-invariant quantity.
The Schwinger-Dyson  equation  approach does not in general
guarantee that the gauge-invariant quantity calculated in
one gauge coincides with that calculated in other gauges.
\par
Even in the approach of the effective potential, a
special gauge is usually taken such as the Landau gauge,
see \cite{SW92}. Note that if we are allowed to change the
order of calculating the graph or equivalently performing the
integration over the internal momenta, we get (see
Fig.~3(b))
\begin{eqnarray}
\int {d^3k \over (2\pi)^3}
D_{\mu \nu}^{(n)}(k) \Pi_{\mu \nu}^{(m)}(k)
=  \int {d^3p \over (2\pi)^3}
S^{(0)}(p) \Sigma^{(1)}(p).
\end{eqnarray}
In this case, even if we start from the second expression, we
should obtain the same result as long as it is finite.
However, it is divergent and hence we need to regularize it.
The naive cutoff procedure usually taken in the calculation
of the effective potential may break the gauge invariance
in the sense that this reordering is not allowed and
hence the result may be not gauge-covariant. Therefore,
if we start from the gauge-noninvariant quantity
$\Sigma(p)$, it is quite difficult to recover the gauge
covariance after the calculation.

\section{Order parameter and vacuum polarization}
\setcounter{equation}{0}

We introduce the projection operators
\begin{equation}
 \chi_{\pm} := {1 \over 2} (1+\tau),
 \quad
 \tau  = \pmatrix{  I  & 0      \cr
               0  & -I   \cr} ,
\end{equation}
which satisfy the properties: $\chi_{\pm}^2=\chi_{\pm}$,
$\chi_{+}\chi_{-}=0$ and $\chi_{+}+\chi_{-}=1$.
Then the source term~(\ref{source}) is rewritten as
\begin{eqnarray}
 {\cal L}_J = J_+ \bar \Psi^\alpha(x) \chi_+ \Psi^\alpha(x)
 + J_- \bar \Psi^\alpha(x) \chi_- \Psi^\alpha(x),
\end{eqnarray}
where we have defined
\begin{equation}
  J_{\pm} := J_e \pm J_o,
\end{equation}
or equivalently
\begin{equation}
  J_{e} = {1 \over 2} (J_+ + J_-), \quad
  J_{o} = {1 \over 2} (J_+ - J_-).
\end{equation}

\subsection{free part}
\par
The bare fermion propagator $S^{(0)}(p)$ is also
decomposed as
\footnote{Note that the signature of $J_e, J_o$ is
opposite to that of the bare fermion mass $m_e, m_o$.}
\begin{equation}
 S^{(0)}(p) = {1 \over {\not p}+J_e+ J_o \tau}
 = S_{+}^{(0)}(p) \chi_{+} + S_{-}^{(0)}(p) \chi_{-},
 \label{decomposition}
\end{equation}
where
\begin{equation}
 S_{\pm}^{(0)}(p) := {1 \over {\not p}+J_{\pm}}.
\end{equation}
\par
Then the free parts of
two order parameters are decomposed as
\begin{eqnarray}
 \langle \bar \Psi \Psi \rangle^{(0)}
 &=& \langle \bar \Psi \Psi \rangle_{+}^{(0)}
 + \langle \bar \Psi \Psi \rangle_{-}^{(0)},
 \nonumber\\
 \langle \bar \Psi \tau \Psi \rangle^{(0)}
 &=& \langle \bar \Psi \Psi \rangle_{+}^{(0)}
 - \langle \bar \Psi \Psi \rangle_{-}^{(0)},
\end{eqnarray}
with
\begin{equation}
\langle \bar \Psi \Psi \rangle_{\pm}^{(0)}
=  \int {d^3p \over (2\pi)^3}
\tr[S_{\pm}^{(0)}(p)\chi_{\pm}] .
\end{equation}
By introducing the UV cutoff $\Lambda_f$, we have
\begin{equation}
\langle \bar \Psi \Psi \rangle_{\pm}^{(0)}
=  \int_0^{\Lambda_f}
{dp \over \pi^2} {p^2 J_{\pm}
\over p^2+J_{\pm}^2} .
\label{chipm}
\end{equation}
Here it should be remarked that the quantity
$\langle \bar \Psi \Psi \rangle_{\pm}^{(0)}$ defined by
eq.~(\ref{chipm}) has an ambiguity arising from  the
lower bound of the integration, no infrared cutoff, unless
we specify the signature of
$J_{\pm}$.  In order to define this quantity unambiguously,
we introduce the small and positive infrared cutoff
$\epsilon>0$ and then take the limit
$\epsilon \downarrow 0$.  Thus we obtain
\begin{eqnarray}
\pi^2 \langle \bar \Psi \Psi \rangle_{\pm}^{(0)}
&=&
\lim_{\epsilon \downarrow 0} \int_\epsilon^{\Lambda_f}
 dp  {p^2 J_{\pm} \over p^2+J_{\pm}^2}
\nonumber\\
&=& \lim_{\epsilon \downarrow 0}
\left[ J_{\pm}(\Lambda_f-\epsilon) + J_{\pm}^2 \left(
\arctan {J_{\pm} \over \Lambda_f} - \arctan {J_{\pm} \over
\epsilon} \right) \right]
\nonumber\\
&=& J_{\pm} \Lambda_f  + J_{\pm}^2 \left(
\arctan {J_{\pm} \over \Lambda_f} -   {\pi \over 2}
sgn(J_{\pm}) \right) ,
\end{eqnarray}
where $sgn(J_{\pm})$ denotes the signature of $J_{\pm}$.
Therefore we have
\begin{eqnarray}
{\langle \bar \Psi \Psi \rangle_{\pm}^{(0)} \over
\Lambda_f^2}  =  {1 \over \pi^2} \left[ {J_{\pm} \over
\Lambda_f}  -  sgn(J_{\pm}) {\pi \over 2}  {J_{\pm}^2
\over \Lambda_f^2} + {J_{\pm}^3 \over \Lambda_f^3}
+ {\cal O}(J_{\pm}^4) \right].
\end{eqnarray}
\par
In what follows, we assume that $J_{+}$ is always
positive, $J_{+}>0$, without loss of generality.
Then, depending on the signature of $J_{-}$, there are
two possibilities:
$J_e>J_o$ for $J_{-}>0$ and  $J_e<J_o$ for $J_{-}<0$.
Here it should be remarked that the signature of $J_e$ and
$J_o$ are not   specified.
\par
First we consider the case (I) in which $J_{+}, J_{-}>0$,
i.e. $J_e>J_o$.  Thus we have
\begin{eqnarray}
 {\langle \bar \Psi \Psi \rangle^{(0)} \over \Lambda_f^2}
 &=& {1 \over \pi^2} \left[ {J_+ + J_- \over \Lambda_f}  -
{\pi \over 2}  {J_+^2+J_-^2 \over \Lambda_f^2}  + {\cal
O}(J^3) \right]
\nonumber\\
 &=& {2 \over \pi^2} \left[ {J_e \over \Lambda_f}
 -  {\pi \over 2}  {J_e^2+J_o^2 \over \Lambda_f^2}
 + {\cal O}(J^3)
\right],
 \nonumber\\
 {\langle \bar \Psi \tau \Psi \rangle^{(0)} \over
 \Lambda_f^2}
 &=& {1 \over \pi^2} \left[ {J_+ - J_- \over \Lambda_f}  -
{\pi \over 2}  {J_+^2 - J_-^2 \over \Lambda_f^2} + {\cal
O}(J^3) \right]
\nonumber\\ &=& {2 \over \pi^2} \left[ {J_o \over
\Lambda_f}  -  {\pi
\over 2}  {2J_e J_o \over \Lambda_f^2}  + {\cal O}(J^3)
\right].
\end{eqnarray}
\par
In the second case (II) where $J_{+}>0, J_{-}<0$, i.e.
$J_o>J_e$.  we obtain
\begin{eqnarray}
 {\langle \bar \Psi \Psi \rangle^{(0)} \over \Lambda_f^2}
 &=& {1 \over \pi^2} \left[ {J_+ + J_- \over \Lambda_f}  -
{\pi \over 2}  {J_+^2-J_-^2 \over \Lambda_f^2}
+ {\cal O}(J^3) \right]
\nonumber\\
 &=& {2 \over \pi^2} \left[ {J_e \over \Lambda_f}  -  {\pi
\over 2}  {2J_e J_o \over \Lambda_f^2} + {\cal O}(J^3)
\right],
 \nonumber\\
 {\langle \bar \Psi \tau \Psi \rangle^{(0)} \over
 \Lambda_f^2}
 &=& {1 \over \pi^2} \left[ {J_+ - J_- \over \Lambda_f}  -
{\pi \over 2}  {J_+^2 + J_-^2 \over \Lambda_f^2}
+ {\cal O}(J^3) \right]
\nonumber\\
 &=& {2 \over \pi^2} \left[ {J_o \over \Lambda_f}  -  {\pi
\over 2}  {J_e^2+ J_o^2 \over \Lambda_f^2} + {\cal O}(J^3)
\right].
\end{eqnarray}

\subsection{vacuum polarization}
\par
By using the decomposition (\ref{decomposition}) for
$S^{(0)}(p)$, the vacuum polarization tensor at one-loop is
also decomposed as
\begin{eqnarray}
 \Pi_{\mu\nu}^{(1)}(k)
 &\equiv& - N_f e^2 \int {d^3p \over (2\pi)^3}
 \tr[\gamma_\mu S^{(0)}(p) \gamma_\nu S^{(0)}(p+k)]
 \nonumber\\
 &=& N_f[\Pi_{\mu\nu}^{(1)}(k;J_{+})
    + \Pi_{\mu\nu}^{(1)}(k;J_{-})] ,
\end{eqnarray}
where we have defined
\begin{equation}
 \Pi_{\mu\nu}^{(1)}(k;J_{\pm})
 := -e^2 \int {d^3p \over (2\pi)^3}  \tr[\gamma_\mu
S^{(0)}_{\pm}(p)\gamma_\nu S^{(0)}_{\pm}(p+k)
\chi_{\pm}] .
\end{equation}
For the one-loop vacuum polarization tensor with the
tensor structure:
\begin{eqnarray}
\Pi_{\mu \nu}^{(1)}(k)
=   \left[ \delta_{\mu \nu}-{k_\mu k_\nu \over k^2} \right]
\Pi_T^{(1)}(k)
 + \epsilon_{\mu\nu\rho}
{k_\rho \over \sqrt{k^2}} \Pi_O^{(1)}(k),
\end{eqnarray}
we find
\begin{eqnarray}
\Pi_T^{(1)}(k)  &=&
N_f[\Pi_T^{(1)}(k;J_{+})+\Pi_T^{(1)}(k;J_{-})],
\nonumber\\
\Pi_O^{(1)}(k)  &=&
N_f[\Pi_O^{(1)}(k;J_{+})-\Pi_O^{(1)}(k;J_{-})] ,
\end{eqnarray}
since the tensor $\Pi_{\mu\nu}^{(1)}(k;J_{\pm})$ should
have the same tensor structure as
$\Pi_{\mu\nu}^{(1)}(k)$.
Note that there exists a minus sign in $\Pi_O^{(1)}(k)$.
Actual calculation shows (see Appendix C) that
\begin{eqnarray}
\Pi_T^{(1)}(k;J) &=&  - {e^2 \over 8\pi} \left[ 2|J| +
{k^2-4J^2 \over k}
\arctan {k \over 2|J|} \right],
\end{eqnarray}
and
\begin{eqnarray}
\Pi_O^{(1)}(k;J) &=&   {e^2 \over 2\pi} J
\arctan {k \over 2|J|}.
\end{eqnarray}
It should be
remarked that
$\Pi_T^{(1)}(k;J)$ is an even function of $J$ and
$\Pi_O^{(1)}(k;J)$ is odd in $J$. Therefore,
if $J_e=0$, then $J_\pm = \pm J_o$ and
\begin{eqnarray}
\Pi_T^{(1)}(k) = 2N_f \Pi_T^{(1)}(k;J_o),
\  \Pi_O^{(1)}(k) = 2N_f \Pi_O^{(1)}(k;J_o).
\end{eqnarray}
On the other hand, if $J_o=0$, then $J_+=J_-=J_e$ and
\begin{eqnarray}
\Pi_T^{(1)}(k) = 2N_f \Pi_T^{(1)}(k;J_e),
\  \Pi_O^{(1)}(k) = 0.
\end{eqnarray}
\par
The usual criterion for generation of the induced
Chern-Simons term  is given by $\Pi_O^{(1)}(k \rightarrow
0)\not=0$.  The induced Chern-Simons  term is generated when
$J_+>0, J_-<0 (J_e<J_o)$, since
\begin{eqnarray}
\Pi_O^{(1)}(k \rightarrow 0)
&=& {N_fe^2 \over 4\pi}[sgn(J_{+})-sgn(J_{-})]
\nonumber\\ &=&
{N_fe^2 \over 4\pi}[sgn(J_{e}+J_{o})-sgn(J_{e}-J_{o})]
\nonumber\\ &=&
\cases{0   & $(J_e>J_o; J_+, J_->0)$ \cr
{N_fe^2 \over 2\pi} & $(J_e<J_o; J_+>0, J_-<0)$ \cr}.
\end{eqnarray}

\section{Order parameter as a function of  source}
\setcounter{equation}{0}
\par
We introduce new mass parameters (or sources) $J_1, J_2$
corresponding to two 2-component fermions, $\psi_1, \psi_2 $:
\begin{eqnarray}
 J_1 = J_{+}, \quad J_2 = - J_{-},
\end{eqnarray}
so that the source term~(\ref{source}) is rewritten as
\begin{eqnarray}
  {\cal L}_J = J_1 \bar \psi_1(x) \psi_1(x)
  + J_2 \bar \psi_2(x) \psi_2(x) ,
\end{eqnarray}
where
\begin{eqnarray}
  J_{e}
  = {1 \over 2} (J_1 - J_2),
\quad
  J_{o}
  = {1 \over 2} (J_1 + J_2),
\end{eqnarray}
or
\begin{eqnarray}
  J_{1}   = J_e + J_o,
\quad
  J_{2}   = - J_e + J_o.
\end{eqnarray}
\par
Similarly in the preceding section, the free parts of the
order parameter are decomposed as
\begin{eqnarray}
 \langle \bar \Psi \Psi \rangle^{(0)}
 &=& \langle \bar \psi_1 \psi_1 \rangle^{(0)}
 - \langle \bar \psi_2 \psi_2 \rangle^{(0)} ,
 \nonumber\\
 \langle \bar \Psi \tau \Psi \rangle^{(0)}
 &=& \langle \bar \psi_1 \psi_1 \rangle^{(0)}
 + \langle \bar \psi_2 \psi_2 \rangle^{(0)} ,
\end{eqnarray}
where we have defined:
\begin{equation}
\langle \bar \psi_{a} \psi_{a} \rangle^{(0)}
=  \int_0^{\Lambda_f}
{dp \over \pi^2} {p^2 J_{a}
\over p^2+J_{a}^2}
\quad (a = 1, 2).
\label{freepart}
\end{equation}

\par
We apply the above decomposition also to the interaction
part.
By taking into account eq.~(\ref{formula}) and
\begin{eqnarray}
 {\partial \over \partial J_e}
 = {\partial \over \partial J_{1}}
   - {\partial \over \partial J_{2}},
\quad {\partial \over \partial J_o}
 = {\partial \over \partial J_{1}}
   + {\partial \over \partial J_{2}},
\end{eqnarray}
the two order parameters are rewritten in the form:
\begin{eqnarray}
 \langle \bar \Psi \Psi \rangle
 &=&  \langle \bar \psi_1 \psi_1 \rangle
 - \langle \bar \psi_2 \psi_2 \rangle ,
 \nonumber\\
 \langle \bar \Psi \tau \Psi \rangle
 &=& \langle \bar \psi_1 \psi_1 \rangle
 + \langle \bar \psi_2 \psi_2 \rangle ,
\end{eqnarray}
where
\begin{eqnarray}
&& \langle \bar \psi_{a} \psi_{a} \rangle
\nonumber\\
&=& \langle \bar \psi_{a} \psi_{a} \rangle^{(0)}
 + {1 \over 2}  N_f^{-1}
\int {d^3k \over (2\pi)^3} D_{\mu \nu}^{(1)}(k)
{\partial \over \partial J_a} \Pi_{\mu \nu}^{(1)}(k),
\nonumber\\
&=& \langle \bar \psi_{a} \psi_{a} \rangle_a^{(0)}
 + N_f^{-1} \int  {d^3k \over (2\pi)^3}
 \{ [\beta k^2-\Pi_T^{(1)}(k)]^2
+[\theta \sqrt{k^2}-\Pi_O^{(1)}(k)]^2 \}^{-1}
\nonumber\\
&& \times \left\{ [\beta k^2-\Pi_T^{(1)}(k)]
{\partial \Pi_T^{(1)}(k) \over \partial J_a}
 + [\theta \sqrt{k^2}-\Pi_O^{(1)}(k)]
{\partial \Pi_O^{(1)}(k) \over \partial J_a}
\right\}
\nonumber\\ &&
+ {\cal O}\left( \left( 1/N_f \right)^2 \right) .
\label{op0}
\end{eqnarray}
\par
Now we study the dependence of the integrand of
Eq.~(\ref{op0}) on the source.
First of all, it is shown by the same procedure as in the
previous section that the free part (\ref{freepart}) reads
\begin{eqnarray}
{\langle \bar \psi_{a} \psi_{a} \rangle^{(0)} \over
\Lambda_f^2}
=  {1 \over \pi^2} \left[ {J_{a} \over \Lambda_f}
-  \sigma_a {\pi \over 2}  {J_{a}^2 \over \Lambda_f^2}
+ {\cal O}(J_{a}^3) \right] ,
\end{eqnarray}
where
$\sigma_a$ denotes the signature of $J_a$, i.e.,
$
\sigma_a = {J_a \over |J_a|}.
$\par
To study the dependence of the interaction part on the
source,  we observe that the one-loop calculation in
Appendix C leads to
\begin{eqnarray}
\Pi_T^{(1)}(k)   &=&  \alpha k \left[ -1 + 2
\sum_a {J_a^2 \over k^2} - {32 \over 3\pi}
\sum_a {\sigma_a J_a^3 \over k^3} \right]
+ {\cal O}\left( J_a^4 \right),
\nonumber\\
\Pi_O^{(1)}(k)   &=& 2\alpha  \left[ \sum_a J_a
- {4 \over \pi} \sum_a {\sigma_a J_a^2 \over k} \right]
   + {\cal O}\left( J_a^4 \right),
\label{vp(0)}
\end{eqnarray}
and
\begin{eqnarray}
 {\partial \Pi_T^{(1)}(k) \over \partial J_a}
   &=&  4\alpha \left[ {J_a \over k}
   - \sigma_a {8 \over \pi} {J_a^2 \over k^2}  \right]
        + {\cal O}(J_{a}^4)  ,
\nonumber\\
 {\partial \Pi_O^{(1)}(k) \over \partial J_a}
   &=& 2\alpha \left[ 1
   - \sigma_a {8 \over \pi} {J_a \over k}  \right]
   + {\cal O}(J_{a}^3) ,
   \label{dvp(0)}
\end{eqnarray}
where  a parameter $\alpha$ is defined by
\begin{eqnarray}
\alpha := {N_f e^2 \over 8},
\end{eqnarray}
which has the dimension of mass in three-dimensions.

\subsection{case (I)}
\par
First we consider the case of $J_1>0, J_2<0$
$(J_+ > 0, J_- > 0)$.
The numerator of the integrand in eq.~(\ref{op0}) has the
following dependence on the source $J_a$:
\begin{eqnarray}
&& [\beta k^2-\Pi_T^{(1)}(k)]
{\partial \Pi_T^{(1)}(k) \over \partial J_a}
 + [\theta \sqrt{k^2}-\Pi_O^{(1)}(k)]
{\partial \Pi_O^{(1)}(k) \over \partial J_a}
\nonumber\\
&=& P_0 + P_1^a J_a + P_2 \sum_b J_b
+ P_3^a J_a^2
+ P_4 \sum_b \sigma_b J_b^2
+ P_5^a J_a \sum_b J_b
+ {\cal O}(J_{a}^3) ,
\label{numerator}
\end{eqnarray}
where
\begin{eqnarray}
 P_0(k)  &=& 2 \theta \alpha k,
 \nonumber\\
 P_1^a(k)  &=& - \sigma_a {16 \over \pi} \theta \alpha
 + 4 \alpha (\beta k + \alpha),
 \nonumber\\
 P_2(k)  &=& - 4 \alpha^2 < 0,
 \nonumber\\
 P_3^a(k)  &=& -\sigma_a {32\alpha \over \pi}
 {\beta k + \alpha \over k} = \sigma_a P_3,
 \nonumber\\
 P_4(k)  &=& {16 \over \pi} {\alpha^2 \over k} > 0,
 \nonumber\\
 P_5^a(k)  &=&
 \sigma_a {32 \over \pi}  {\alpha^2 \over k}
 = \sigma_a P_5.
\end{eqnarray}
On the other hand, the denominator of the integrand in
eq.~(\ref{op0}) has
\begin{eqnarray}
&& \{ [\beta k^2-\Pi_T^{(1)}(k)]^2
+[\theta \sqrt{k^2}-\Pi_O^{(1)}(k)]^2 \}^{-1}
\nonumber\\
&=&  [F(k)]^{-1}
\Biggr[ 1 + Q_0(k)  \sum_a J_a
\nonumber\\&&
+ Q_1(k) \sum_a \sigma_a J_a^2
+ Q_2(k) \sum_a J_a^2
+ Q_3(k) \left( \sum_a J_a \right)^2
\Biggr]
+ {\cal O}(J_{a}^3) ,
\label{denominator}
\end{eqnarray}
where
\begin{eqnarray}
 Q_0(k)  &=& {4\theta \alpha k \over F(k)}
 \nonumber\\
 Q_1(k)  &=& - 4\alpha {{4\theta \over \pi} \over F(k)},
 \nonumber\\
 Q_2(k)  &=& 4\alpha {\beta k + \alpha  \over F(k)},
 \nonumber\\
 Q_3(k)  &=& - 4\alpha^2{[(\beta k^2 + \alpha k)^2 +
(\theta k)^2 ] - 4 \theta^2 k^2 \over F(k)^2},
\end{eqnarray}
and
\begin{eqnarray}
 F(k) := (\beta k^2 + \alpha k)^2 + (\theta k)^2 .
\end{eqnarray}
\par
Substituting eq.~(\ref{numerator}) and
eq.~(\ref{denominator}) into eq.~(\ref{op0}), we obtain
for
$a=1,2$
\begin{eqnarray}
\langle \bar \psi_a \psi_a \rangle
&=& \langle \bar \psi_a \psi_a \rangle^{(0)}
 + N_f^{-1} \int  {d^3k \over (2\pi)^3}
 F(k)^{-1}
\Biggr[
P_0 + P_1^a  J_a + A \sum_b J_b
\nonumber\\&&
+ P_3^a J_a^2 + B \sum_b \sigma_b J_b^2
+ C^a J_a \sum_b J_b
+ D \sum_b J_b^2 + E (\sum_b J_b)^2
\Biggr]
\nonumber\\&&
+ {\cal O}\left( J^3 \right) ,
\end{eqnarray}
where
\begin{eqnarray}
 A(k)  &=&  P_0(k) Q_0(k) + P_2(k),
 \nonumber\\
 B(k)  &=&  P_0(k) Q_1(k) + P_4(k) ,
 \nonumber\\
 C^a(k)  &=&  P_1^a(k) Q_0(k) + P_5^a(k),
 \nonumber\\
 D(k)  &=&  P_0(k) Q_2(k) ,
 \nonumber\\
 E(k)  &=&  P_0(k) Q_3(k) + P_2 Q_0(k) .
 \label{coeffi}
\end{eqnarray}
Thus we obtain the order parameter in the form:
\begin{eqnarray}
 \varphi_1  &:=&
 \langle \bar \psi_1^\alpha \psi_1^\alpha \rangle/N_f
 =\varphi_\theta + u_1 J_1 + v J_2
 + s_1 (J_1)^2 + t_1 (J_2)^2 + r_1 J_1 J_2
 + {\cal O}\left( J^3 \right),
 \nonumber\\
 \varphi_2  &:=&
 \langle \bar \psi_2^\alpha \psi_2^\alpha \rangle/N_f
= \varphi_\theta + v J_1 + u_2 J_2
 + t_2 (J_1)^2 + s_2 (J_2)^2 + r_2 J_1 J_2
 + {\cal O}\left( J^3 \right),
  \label{order10}
\end{eqnarray}
where
\begin{eqnarray}
 u_a  &=&  \int_0^{\Lambda_f}
{dp \over \pi^2}
+ {1 \over N_f}
\int {d^3k \over (2\pi)^3}
 [P_1^a(k)+A(k)]/F(k),
 \nonumber\\
 v  &=& {1 \over N_f}
\int  {d^3k \over (2\pi)^3}
A(k)/F(k),
 \nonumber\\
 s_a  &=& {1 \over N_f}
 \int  {d^3k \over (2\pi)^3}
 [P_3^a(k)+\sigma_a B(k)+C^a(k)+D(k)+E(k)]/F(k),
 \nonumber\\
 t_a  &=&  {1 \over N_f}
 \int   {d^3k \over (2\pi)^3}
 [-\sigma_a B(k)+D(k)+E(k)]/F(k),
 \nonumber\\
r_a  &=& {1 \over N_f}
\int   {d^3k \over (2\pi)^3}
 [C^a(k) + 2 E(k)]/F(k),
\end{eqnarray}
and
\begin{eqnarray}
 \varphi_\theta
 &=& {1 \over N_f} \int {d^3k \over (2\pi)^3}
 P_0(k)/F(k).
\end{eqnarray}
It is easy to check that
\begin{eqnarray}
 r_1 = 2t_2, \quad r_2 = 2t_1,
\end{eqnarray}
since
$P_5=2P_4$ and
$P_1^a Q_0 = 2P_0(Q_2+\sigma_a Q_1)$.
\par
The parity even and odd order parameters are written as
\begin{eqnarray}
 \varphi_e &:=&
 \langle \bar \Psi^\alpha  \Psi^\alpha \rangle/N_f
= \varphi_{+} + \varphi_{-}
 = \varphi_{1} - \varphi_{2},
 \nonumber\\
 \varphi_o &:=&
 \langle \bar \Psi^\alpha \tau \Psi^\alpha \rangle/N_f
 = \varphi_{+} - \varphi_{-}
 = \varphi_{1} + \varphi_{2}.
\end{eqnarray}
They obey
\begin{eqnarray}
 \varphi_e &=&   (u_1+u_2-2v) J_e  + (u_1-u_2) J_o
\nonumber\\&&
  + [s_1-s_2+3(t_1-t_2)] J_e^2
 + [s_1-s_2-(t_1-t_2)] J_o^2
\nonumber\\&&
 + 2[s_1+s_2-(t_1+t_2)] J_e J_o
 + {\cal O}(J^3),
 \nonumber\\
 \varphi_o &=&  2 \varphi_\theta + (u_1-u_2) J_e
 + (u_1+u_2+2v) J_o
\nonumber\\&&
 + [s_1+s_2-(t_1+t_2)] J_e^2
  + [s_1+s_2+3(t_1+t_2)] J_o^2
\nonumber\\&&
 + 2[s_1-s_2-(t_1-t_2)] J_e  J_o
 + {\cal O}(J^3) .
 \label{two-op}
\end{eqnarray}
\par
In order to study the small $\theta$ case,
we first observe that
\begin{eqnarray}
 A(k)  &=&   P_2(k) + {\cal O}(\theta^2) < 0,
 \nonumber\\
 B(k)  &=&   P_4(k) + {\cal O}(\theta^2) > 0,
 \nonumber\\
 C^a(k)  &=&  P_5^a(k)
 + {16 \alpha^2 k(\beta k+\alpha) \over F_0(k)} \theta
 + {\cal O}(\theta^2) ,
 \nonumber\\
 D(k)  &=&
 {8 \alpha^2 k(\beta k+\alpha) \over F_0(k)} \theta
 + {\cal O}(\theta^2),
 \nonumber\\
 E(k)  &=&  - {24\alpha^3 k \over F_0(k)} \theta
 + {\cal O}(\theta^2) ,
\end{eqnarray}
where
\begin{eqnarray}
 F_0(k) := (\beta k^2 + \alpha k)^2 .
\end{eqnarray}
Note that when $\theta = 0$,
$P_0=0$, $Q_0=0$ , $Q_1=0$ and $P_1^1=P_1^2$.
Therefore, the functions $Q_a (a=1,2,3)$ do not contribute
to the  order parameter.
\par
Thus the coefficient is calculated
up to ${\cal O}(1/N_f)$ and ${\cal O}(\theta)$:
\begin{eqnarray}
 u_a  &=&  \int_0^{\Lambda_f} {dp \over \pi^2}
 + {1 \over N_f}
\int  {d^3k \over (2\pi)^3}
 {4 \alpha \beta k \over F_0(k)}
 \nonumber\\&&
 - {\theta \over N_f} \sigma_a
\int {d^3k \over (2\pi)^3}
 {{16 \over \pi}  \alpha \over F_0(k)}
 + {\cal O}\left( \theta^2/N_f^2 \right),
 \nonumber\\
 v  &=& - {1 \over N_f}
 \int  {d^3k \over (2\pi)^3}
{4\alpha^2 \over F_0(k)}
 + {\cal O}\left( \theta^2/N_f^2 \right),
 \nonumber\\
 s_a  &=& {1 \over N_f}
 \int  {d^3k \over (2\pi)^3}
 {\sigma_a [P_3+P_4(k)+P_5(k)] \over F_0(k)}
 \nonumber\\&&
 + {\theta  \over N_f} \int {d^3k \over (2\pi)^3}
 {24 \alpha^2  \beta k \over F_0(k)^2}
 + {\cal O}\left( \theta^2/N_f^2 \right),
 \nonumber\\
 t_a  &=&  - {1 \over N_f}
 \int   {d^3k \over (2\pi)^3}
 {\sigma_a P_4(k) \over F_0(k)}
 \nonumber\\&&
 + {\theta  \over N_f} \int {d^3k \over (2\pi)^3}
 {8 \alpha^2 k(\beta k-2\alpha) \over F_0(k)^2}
 + {\cal O}\left( \theta^2/N_f^2 \right),
 \label{coeff}
\end{eqnarray}
and
\begin{eqnarray}
 \varphi_\theta
 &=& {\theta \over N_f}
 \int  {d^3k \over (2\pi)^3}
 {2\alpha k \over F_0(k)}
  + {\cal O}\left( \theta^3/N_f^2 \right).
\end{eqnarray}
It is easy to see that
\begin{eqnarray}
 u_1 - u_2 &=&  {\cal O}\left( \theta/N_f \right),
\quad
 u_1 + u_2
 =  (u_1 + u_2)_{\theta=0}
 + {\cal O}\left( \theta^2/N_f^2 \right),
 \nonumber\\
 s_1 + s_2 &=&  {\cal O}\left( \theta/N_f \right),
\quad
 s_1 - s_2 =  (s_1 - s_2)_{\theta=0}
 + {\cal O}\left( \theta^2/N_f^2 \right),
 \nonumber\\
 t_1 + t_2 &=&  {\cal O}\left( \theta/N_f \right),
\quad
 t_1 - t_2 = (t_1 - t_2)_{\theta=0}
 + {\cal O}\left( \theta^2/N_f^2 \right).
 \label{estimate}
\end{eqnarray}
Hence, for example, we obtain
$
 u_1 + u_2 \pm 2v
 = 2(u_{\theta=0} \pm v_{\theta=0})
 + {\cal O}\left( \theta^2/N_f^2 \right).
$
\par
Finally, we consider the limit $\theta \rightarrow
0$, i.e., the Chern-Simons term is absent.
In this limit, we find
$s_1 + s_2 \rightarrow 0$ and $t_1+t_2 \rightarrow 0$,
and $u_1, u_2$ approach to the same limit:
$u_1, u_2 \rightarrow u$.
Thus, in the limit $\theta \rightarrow 0 $,
two order parameters obey
\begin{eqnarray}
 \varphi_e &=&  2(u-v) J_e + [s_1-s_2+3(t_1-t_2)] J_e^2 +
[s_1-s_2-(t_1-t_2)] J_o^2
 + {\cal O}(J^3),
 \nonumber\\
 \varphi_o &=&   2(u+v) J_o
 + 2[s_1-s_2-(t_1-t_2)] J_e J_o
 + {\cal O}(J^3),
 \label{order1}
\end{eqnarray}
where
\begin{eqnarray}
  u &=&  \int_0^{\Lambda_f} {dp \over \pi^2}
 + {1 \over N_f}
\int  {d^3k \over (2\pi)^3}
 {4 \alpha \beta k \over F_0(k)}
 + {\cal O}\left({1 \over N_f^2}\right),
 \nonumber\\
 v  &=&   {-1 \over N_f}
 \int  {d^3k \over (2\pi)^3}
{4\alpha^2 \over F_0(k)}
 + {\cal O}\left({1 \over N_f^2}\right),
 \nonumber\\
 s_a  &=& {1 \over N_f}
 \int  {d^3k \over (2\pi)^3}
 {\sigma_a \over F_0(k)}
 \left({16 \alpha \over \pi}\right)
 {-2\beta k + \alpha \over k}
  + {\cal O}\left({1 \over N_f^2}\right),
 \nonumber\\
 t_a  &=&   {-1 \over N_f}
 \int  {d^3k \over (2\pi)^3}
 {\sigma_a  \over F_0(k)}
 \left({16 \over \pi}\right) {\alpha^2 \over k}
  + {\cal O}\left({1 \over N_f^2}\right).
\end{eqnarray}
For example, if we introduce another cutoff $\Lambda_p$, we
get
\begin{eqnarray}
 2(u-v)  =  {2\Lambda_f \over \pi^2}
 \left[ 1  + {K_1 \over N_f} \right],
\end{eqnarray}
where
\begin{eqnarray}
 K_1 &=&
{\pi^2 \over \Lambda_f}
\int^{\Lambda_p} {d^3k \over (2\pi)^3}
 {4\alpha(\beta k+\alpha) \over
 (\beta k^2 + \alpha k)^2}
 \nonumber\\
 &=&   2 {\Lambda_p \over \Lambda_f}
 \left( {\beta \Lambda_p \over \alpha} \right)^{-1}
 \ln \left( 1 + {\beta \Lambda_p \over \alpha} \right) .
 \label{K1}
\end{eqnarray}

\subsection{case (II)}
\par
Next we consider the case of
$J_1>0, J_2>0$ $(J_+>0, J_-<0)$.
\par
In this case we take $\sigma_a=1$ for $a=1,2$ in
Eq.~(\ref{vp(0)}) and Eq.~(\ref{dvp(0)}).
Then we obtain
\begin{eqnarray}
 \varphi_1  &=&
 \varphi_\theta + u J_1 + v J_2
 + s (J_1)^2 + t (J_2)^2 + 2t J_1 J_2
 + {\cal O}\left( J^3 \right),
 \nonumber\\
 \varphi_2  &=&
\varphi_\theta + v J_1 + u J_2
 + t (J_1)^2 + s (J_2)^2 + 2t J_1 J_2
 + {\cal O}\left( J^3 \right),
 \label{order20}
\end{eqnarray}
where the coefficients $u,v,s,t$ are given by
Eq.~(\ref{coeff}) with $\sigma_a \equiv 1$.
Hence the order parameter
is written in the form:
\begin{eqnarray}
 \varphi_e &=&  2(u-v) J_e + 4(s-t) J_e J_o
 + {\cal O}(J^3),
 \nonumber\\
 \varphi_o &=&  2 \varphi_\theta + 2(u+v) J_o
 + 2(s-t) J_e^2 + 2(s+3t) J_o^2
 + {\cal O}(J^3).
 \label{order2}
\end{eqnarray}
When $\theta=0$, for example, we get
\begin{eqnarray}
 2(u+v)  =  {2\Lambda_f \over \pi^2}
 \left[ 1  + {L_1 \over N_f} \right],
\end{eqnarray}
where
\begin{eqnarray}
 L_1  &=&
{\pi^2 \over \Lambda_f}
\int^{\Lambda_p} {d^3k \over (2\pi)^3}
 {4\alpha(\beta k-\alpha) \over
 (\beta k^2 + \alpha k)^2}
 \nonumber\\
 &=&   2 {\Lambda_p \over \Lambda_f}
 \left( {\beta \Lambda_p \over \alpha} \right)^{-1}
 \left[
 \ln \left( 1 + {\beta \Lambda_p \over \alpha} \right)
 - 2 {{\beta \Lambda_p \over \alpha} \over
 1 + {\beta \Lambda_p \over \alpha}} \right] .
 \label{L1}
\end{eqnarray}

\section{Inversion method}
\setcounter{equation}{0}
In this section we restrict our consideration to the case
in which the bare Chern-Simons term is absent.
The effect of the bare Chern-Simons term will be
investigated in section 8.
In this paper we use the inversion method to obtain the
effective potential and find the non-trivial solution as
its stationary point.  The inversion method was
proposed by Fukuda \cite{Fukuda88} and was applied to gauge
theories: for example, QCD4 \cite{Fukuda88}, the strong
coupling QED4 \cite{UKF90,Yokojima95}, the gauged Yukawa
model
\cite{Kondo93,KITE94} and the Thirring model (reformulated
as a gauge theory) \cite{Kondo95th}.

\subsection{general strategy}
Let $g$ denote an expansion parameter in a certain scheme
of the expansion.  For example, $g$ may be the coupling
constant $e^2$ in the usual perturbation theory or the
expansion parameter $1/N_f$ in the $1/N_f$ expansion.
In order to calculate the order parameter
$\langle \bar \Psi \Psi \rangle$ and/or
$\langle \bar \Psi \tau \Psi \rangle$
we introduce the source term
$
{\cal L}_J = J_e \bar \Psi^\alpha(x) \Psi^\alpha(x)
+ J_o \bar \Psi^\alpha(x) \tau \Psi^\alpha(x)
$
into the original Lagrangian.
We define the dimensionless
\footnote{
Here we have adopted the UV cutoff $\Lambda_f$ to make the
order parameter  dimensionless.  However this is not a
unique choice.  We can take other quantities which play
the same role, e.g., $\alpha := N_f e^2$.
The result on the existence of the phase transition is
unchanged irrespective of what quantity we may choose to
define the dimensionless order parameter, as already
pointed out in
\cite{Kondo95th}. }
order parameter $\phi$ and
$\chi$ by
\begin{eqnarray}
\phi &:=& {\pi^2 \over 2\Lambda_f^2}
 {\langle \bar \Psi^\alpha \Psi^\alpha \rangle \over N_f},
 \nonumber\\
 \chi &:=& {\pi^2 \over 2\Lambda_f^2}
 {\langle \bar \Psi^\alpha \tau \Psi^\alpha \rangle \over
N_f},
\label{dimensionlessop}
\end{eqnarray}
with an appropriate factor which simplifies the following
calculations.
In a certain expansion scheme, suppose that
the order parameter is calculated in the power series of
$g$:
\begin{eqnarray}
\phi &=& h_e[J_e, J_o]
= \sum_{n=0}^\infty g^n \phi_n(J_e, J_o),
\nonumber\\
\chi &=& h_o[J_e, J_o]
= \sum_{n=0}^\infty g^n \chi_n(J_e, J_o).
\end{eqnarray}
Then we invert these equations with respect to the source:
\begin{eqnarray}
J_e &=& f_e[\phi,\chi]
= \sum_{m=0}^\infty g^m f_m^e(\phi,\chi),
\nonumber\\
J_o &=& f_o[\phi,\chi]
= \sum_{m=0}^\infty g^m f_m^o(\phi, \chi).
\label{invseries}
\end{eqnarray}
We call this procedure the inversion.
Here the coefficient functions
$f_m^e(\phi, \chi)$ and $f_m^o(\phi, \chi)$ are
determined from the requirement that the following
consistency condition should hold:
\begin{eqnarray}
\phi &=& h_e[f_e[\phi,\chi], f_o[\phi,\chi]],
\nonumber\\
\chi &=& h_o[f_e[\phi,\chi], f_o[\phi,\chi]],
\end{eqnarray}
and are written by using
$\phi_n(J_e, J_o)$ and $\chi_n(J_e, J_o)$.
In the inverted series eq.~(\ref{invseries}), we look for
the non-trivial solution for $\phi$ and $\chi$ in the
limit, $J_e \rightarrow 0$ and $J_o  \rightarrow 0$,
besides the trivial solution:
$\phi = 0$ and $\chi = 0$.
This is the inversion method.  Thus we can get the
non-perturbative solution
$\phi \not= 0$ or $\chi \not= 0$,
if it exists at all, by the inversion method
\cite{Fukuda88}.

\subsection{case (I)}
\par
In the case of $J_{1}>0, J_{2}<0$ ($J_e>J_o$),
eq.~(\ref{order1}) shows that, up to the leading order of
$g=1/N_f$,
$\phi$ and $\chi$ are written as functions of
two sources $J_e, J_o$:
\begin{eqnarray}
 \phi &:=& K \tilde J_e - (P\tilde J_e^2 + Q\tilde J_o^2)
 +  {\cal O}(J^3),
\nonumber\\
 \chi &:=& L \tilde J_o - R(2\tilde J_e \tilde J_o)
 +  {\cal O}(J^3),
 \label{expansion1}
\end{eqnarray}
where we have defined the dimensionless source:
\begin{eqnarray}
\tilde J_{e,o} := {J_{e,o} \over \Lambda_f}.
\end{eqnarray}
Here we find that the cross term $J_e J_o$ does not
appear in
$\phi$ and that $J_e^2, J_o^2$ terms do not appear in
$\chi$.  In what follows ${\cal O}(J^3)$ implies
${\cal O}(J_e^3)$, ${\cal O}(J_e J_o^2)$,
${\cal O}(J_e^2 J_o)$ or ${\cal O}(J_o^3)$.
The coefficients
$K, L, P, Q, R$ are written as power series of $g$, e.g.,
$K=\sum_{n=0}^\infty g^n K_n$.
In this paper we pay attention to the first two terms
$K_0, K_1$  up to ${\cal O}(g)$ and similarly for other
coefficients, $L, P, Q, R$.
Our definition of $\phi, \chi$ leads to
\begin{eqnarray}
 K_0 = L_0 = 1, \ P_0 = Q_0 = R_0 = {\pi \over 2},
\end{eqnarray}
as already shown in section 3.
\par
Now we invert the above set of equations with respect to
two sources, $J_e$ and  $J_o$. We assume the inverted form
as
\begin{eqnarray}
\tilde J_e &=& \tau_e \phi + A \phi^2 + B \chi^2
+ C \phi \chi + {\cal O}(\varphi^3),
 \nonumber\\
\tilde J_o &=& \tau_o \chi  + D \phi^2 + E \chi^2
+ F \phi \chi  + {\cal O}(\varphi^3),
\label{inv0}
\end{eqnarray}
where ${\cal O}(\varphi^3)$ denotes ${\cal O}(\phi^3)$,
${\cal O}(\phi^2 \chi)$, ${\cal O}(\phi \chi^2)$ or
${\cal O}(\chi^3)$.
\par
{}From the condition for the inverted series
eq.~(\ref{inv0}) to be consistent with the above equation
eq.~(\ref{expansion1}),  the coefficient is determined.
Hence the inverted series in the case (I) is obtained as
\begin{eqnarray}
\tilde J_e &=& \tau_e \phi + A \phi^2 + B \chi^2
+ {\cal O}(\varphi^3),
 \nonumber\\
\tilde J_o &=& \tau_o \chi
+ F \phi \chi  + {\cal O}(\varphi^3),
\label{inv1}
\end{eqnarray}
where
\begin{eqnarray}
\tau_e &=& K^{-1} ,
\nonumber\\
\tau_o &=& L^{-1} ,
\nonumber\\
 A &=&  P \tau_e^2 K^{-1}  ,
\nonumber\\
 B &=&  Q \tau_o^2 K^{-1}  = Q K^{-1} L^{-2},
 \nonumber\\
 F &=& 2R \tau_e \tau_o L^{-1}  = 2R  K^{-1} L^{-2} ,
\end{eqnarray}
together with
$C = D = E = 0$.
Therefore, up to the leading order of $g$, we obtain
\begin{eqnarray}
\tau_e &=&  1 - K_1 g + {\cal O}(g^2),
\nonumber\\
\tau_o &=&  1 - L_1 g + {\cal O}(g^2),
\nonumber\\  A &=&  P_0 + (P_1-3K_1 P_0)g + {\cal O}(g^2),
\nonumber\\  B &=&  Q_0 + (Q_1-K_1 Q_0 - 2 L_1 Q_0)g +
{\cal O}(g^2),
\nonumber\\  F &=& 2R_0 + 2(R_1-K_1 R_0 - 2 L_1 R_0)g +
{\cal O}(g^2).
\end{eqnarray}

\subsection{case (II)}
\par
Next we consider the case where $J_{1}, J_{2}>0$
$(J_e < J_o)$.
In this case eq.~(\ref{order2}) leads to
\begin{eqnarray}
 \phi &:=& K \tilde J_e  - U(2\tilde J_e \tilde J_o)
 +  {\cal O}(J^3),
\nonumber\\
 \chi &:=& L \tilde J_o  - (S\tilde J_e^2 + T\tilde J_o^2)
 +  {\cal O}(J^3).
 \label{expansion2c}
\end{eqnarray}
Using the consistency condition, we obtain the inverted
equation:
\begin{eqnarray}
\tilde J_e &=& \tau_e \phi  + C \phi \chi
+ {\cal O}(\varphi^3),
 \nonumber\\
\tilde J_o &=& \tau_o \chi  + D \phi^2 + E \chi^2   +
{\cal O}(\varphi^3),
\label{inv2}
\end{eqnarray}
where
\begin{eqnarray}
 \tau_e &=& K^{-1} ,
\nonumber\\
 \tau_o &=& L^{-1} ,
\nonumber\\ C &=&  2U \tau_e \tau_o K^{-1},
\nonumber\\ D &=&  S \tau_e^2 L^{-1},
\nonumber\\ E &=&  T \tau_o^2 L^{-1},
\end{eqnarray}
and $A=B=F$.
Since
\begin{eqnarray}
 K_0 = L_0 = 1, \
 S_0 = T_0 = U_0 = {\pi^2 \over 2},
\end{eqnarray}
we obtain
\begin{eqnarray}
 C_0 = 2D_0 = 2E_0.
\end{eqnarray}

\section{Chiral symmetry breaking in pure QED3}
\setcounter{equation}{0}

In this section we consider the pure QED3 without the bare
Chern-Simons term.
The induced Chern-Simons term $\Pi_O^{(1)}(k)$ is also
neglected in this section as well as the bare Chern-Simons
term.
The Chern-Simons terms are included in the subsequent
sections based on the result of section 4.
This section is an independent one which illustrates the
inversion method in a slightly simplified setting.

\subsection{single order parameter in QED3}
As a special case of the inversion scheme presented in the
previous section,  we introduce the single source term
${\cal L}_J = J_e \bar \Psi(x) \Psi(x)$ and consider a single
order parameter $\phi$ which has the power-series expansion
with respect to the expansion parameter $g$ ($g=e^2$ or
$1/N_f$):
\begin{eqnarray}
\phi &=& h [J_e]
= \sum_{n=0}^\infty g^n \phi_n(J_e).
\end{eqnarray}
After inversion, we will get
\begin{eqnarray}
J_e = f [\phi] =
\sum_{m=0}^\infty g^m f_m (\phi).
\end{eqnarray}
The coefficient function
$f_m(\phi)$ is related to $\phi_n(J_e)$ from the
consistency condition:
\begin{eqnarray}
\phi &=& h [f [\phi]].
\end{eqnarray}
\par
In QED3 without the Chern-Simons term
$(\beta=1, \theta=0)$, it is shown in the following that
$\phi$ obeys
\begin{eqnarray}
 \phi &:=& K \tilde J_e - P \tilde J_e^2
 +  {\cal O}(\tilde J_e^3).
\end{eqnarray}
The inverted equation is obtained as
\begin{eqnarray}
\tilde J_e &=& \tau \phi + A \phi^2  + {\cal O}(\phi^3),
\label{Je}
\end{eqnarray}
where the consistency condition leads to
\begin{eqnarray}
\tau  &=& K^{-1} ,
 \nonumber\\ A &=&  P \tau^2 K^{-1}  .
\end{eqnarray}
Hence, to the leading order of $g$,
\begin{eqnarray}
\tau &=&  1 - K_1 g + {\cal O}(g^2),
\nonumber\\  A &=&   P_0 + (P_1-3K_1 P_0)g + {\cal O}(g^2).
\end{eqnarray}
Even in the limit $J_e \rightarrow 0$, $\phi$ in
eq.~(\ref{Je}) has an non-trivial solution
$\phi = -\tau A^{-1}>0$ when $\tau <0$ (for $A>0$).
Moreover, the effective potential is obtained as
\begin{eqnarray}
 V(\phi) = {\tau \over 2} \phi^2 + {A \over 3} \phi^3,
\end{eqnarray}
from the relation
\begin{eqnarray}
  {\delta V(\phi) \over \delta \phi} = \tilde J_e .
\end{eqnarray}
Hence the non-trivial solution $\phi > 0$ is
energetically more favorable than the trivial one $\phi=0$
when $\tau <0$ (as long as $A>0$). Therefore the chiral
symmetry is spontaneously broken for
$g>g_c=K_1^{-1}$ where $g_c$ is identified with the critical
point.

\subsection{quenched QED3}

In this subsection we consider the chiral symmetry
breaking in the quenched QED3.
Here the term "quenched" implies that all the fermion loop
correction to the photon propagator are neglected. This
situation is realized by taking the limit $N_f \rightarrow
0$ in an appropriate way.  Therefore we estimate the
critical point in the lowest order of the coupling constant
$e$ instead of $1/N_f$.  As shown in section 2, we need to
calculate
\begin{eqnarray}
&& {1 \over 2}
\int {d^3k \over (2\pi)^3} D_{\mu \nu}^{(0)}(k)
{\partial \over \partial J_e} \Pi_{\mu\nu}^{(1)}(k)
\nonumber\\
 &=& {1 \over 4\pi^2} \int_0^{\Lambda_p} k^2 dk
 D_{\mu \nu}^{(0)}(k)
{\partial \over \partial J_e} \Pi_{\mu \nu}^{(1)}(k)
\nonumber\\
 &=& {1 \over 4\pi^2} \int_0^{\Lambda_p}  dk
 {2 \over \beta }
 {\partial \over \partial J_e} \left[ \Pi_T^{(1)}(k)
 \right].
\end{eqnarray}
To the lowest order ${\cal O}(e^2)$, we obtain
\begin{eqnarray}
&& {1 \over 2}
\int {d^3k \over (2\pi)^3} D_{\mu \nu}^{(0)}(k)
{\partial \over \partial J_e} \Pi_{\mu\nu}^{(1)}(k)
\nonumber\\
 &=& {4\alpha \over \pi^2 \beta}\int_\mu^{\Lambda_p} dk
 \left( {J_e \over k} - {8 \over \pi} {J_e^2 \over k^2}
 \right) + {\cal O}(J^4),
\end{eqnarray}
where we have temporally introduced the infrared (IR) cutoff
$\mu$ besides the ultraviolet (UV) cutoff $\Lambda_p$.
For a while, we put $\beta=1$.
In the quenched limit, all the loop
corrections are neglected and hence the photon propagator
reduces to the bare one and the order parameter $\phi$ is
defined by taking the limit
$N_f \rightarrow 0$.
Then we have
\begin{eqnarray}
\phi := \lim_{N_f \rightarrow 0} {\pi^2 \over 2}
{\langle \bar \Psi^\alpha \Psi^\alpha \rangle \over N_f
\Lambda_f^2}
= \tilde J_e \left( 1 + {e^2 \over 8\Lambda_f}
 \ln {\Lambda_p \over \mu} \right)
 - P \tilde J_e^2 + {\cal O}(J^4).
\end{eqnarray}
After inversion up to ${\cal O}(e^2)$, we get
\begin{eqnarray}
\tilde J_e := {J_e \over \Lambda_f}  =  \phi
\left( 1 - {e^2 \over 8\Lambda_f} \ln {\Lambda_p \over \mu}
\right)
+ A \phi^2  + {\cal O}(\phi^4).
\end{eqnarray}
Even when $J_e=0$, there may exist a non-trivial solution
$\phi \not= 0$ besides the trivial one, $\phi = 0$:
\begin{eqnarray}
 \phi = A^{-1}
\left( {e^2 \over 8\Lambda_f} \ln {\Lambda_p \over \mu} - 1
\right) .
\end{eqnarray}
Actually, defining the dimensionless inverse coupling
constant $\beta_e := \Lambda_f/e^2$, we see there is a
non-trivial solution when $\beta_e<\beta_e^c$ for a
critical value:
\begin{eqnarray}
\beta_e^c := {1 \over 8}\ln {\Lambda_p \over \mu} .
\end{eqnarray}
This result however shows that
$\beta_e^c \uparrow \infty$
as $\mu \downarrow 0$.
Naively this may be interpreted as follows.
\footnote{ Note that we obtain
$$
 P_1 = {\pi^2 \over 2} {4\alpha \over \pi^2}
 {8 \over \pi}
 \int_\mu^{\Lambda_p}  dk   {1 \over \beta k^2} ,
$$
which diverges as $\mu \downarrow 0$.
}
In the quenched QED3, the chiral symmetry is spontaneously
broken for arbitrary values of the coupling constant
$e^2\not=0$.  This agrees with the results of the
Schwinger-Dyson equation approach
\cite{KN90} and the Monte Carlo simulation of the lattice
non-compact QED3 \cite{DKK89}.
\par

\subsection{unquenched QED3 in the $1/N_f$ expansion}
Next we consider the problem of chiral symmetry breaking in
unquenched QED3  based on  the scheme of the $1/N_f$
expansion. To the leading order in $1/N_f$ expansion, we wish
to calculate
\begin{eqnarray}
&& {1 \over 2}
\int {d^3k \over (2\pi)^3} D_{\mu \nu}^{(1)}(k)
{\partial \over \partial J_e} \Pi_{\mu \nu}^{(1)}(k)
\nonumber\\
 &=& {1 \over 2\pi^2} \int_0^{\Lambda_p} k^2 dk
  D_{\mu \nu}^{(1)}(k)
{\partial \over \partial J_e} \Pi_{\mu \nu}^{(1)}(k)
\nonumber\\
 &=& {1 \over 2\pi^2} \int_0^{\Lambda_p} k^2 dk   {1 \over
[\beta k^2-\Pi_T^{(1)}(k)]}{\partial \Pi_T^{(1)}(k) \over
\partial J_e}.
\end{eqnarray}
Up to ${\cal O}(J_e^2)$, this reduces to
\begin{eqnarray}
&& {1 \over 2}
\int {d^3k \over (2\pi)^3} D_{\mu \nu}^{(1)}(k)
{\partial \over \partial J_e} \Pi_{\mu\nu}^{(1)}(k)
\nonumber\\
&=& {1 \over 2\pi^2} \int_0^{\Lambda_p} k^2 dk
 {1 \over [\beta k^2+\alpha k]}
 8\alpha \left( {J_e \over k} - {8 \over \pi}{J_e^2 \over
k^2} \right)  + {\cal O}(J_e^3).
\end{eqnarray}
\par
Therefore we obtain
\begin{eqnarray}
\phi := {\pi^2 \over 2}
{\langle \bar \Psi^\alpha \Psi^\alpha \rangle  \over N_f
\Lambda_f^2}
=  \tilde J_e \left( 1 + {K_1 \over  N_f} \right)  - P
\tilde J_e^2 + {\cal O}(J_e^3).
\end{eqnarray}
After inversion, we get
\begin{eqnarray}
\tilde J_e
=  \left( 1 - {N_f^c \over N_f} \right) \phi
 + A \phi^2 + {\cal O}(\phi^3),
\end{eqnarray}
with
\footnote{
Note that we obtain
$$
 P_1 = {\pi^2 \over 2} {4\alpha \over \pi^2}
 {8 \over \pi}
 \int_\mu^{\Lambda_p}  dk   {1 \over k(\beta k+\alpha)} .
$$
Naively this diverges as $\mu \downarrow 0$. The
infrared divergence in $P_1$ can not be removed in this
method.
This term may induce the fluctuation-induced first order
transition as discussed in the final section.
}
\begin{eqnarray}
N_f^c  = K_1 = {\pi^2 \over 2\Lambda_f}
 {4\alpha \over \pi^2}
 \int_0^{\Lambda_p}  dk   {1 \over \beta k+\alpha}
 =  {2\alpha \over \beta\Lambda_f}
 \ln(1+{\beta \Lambda_p \over \alpha}).
\end{eqnarray}
\par
\par
In the leading order of the $1/N_f$ expansion, we have shown
the existence and finiteness of the critical number of
flavors
$N_f^c$, below which the chiral symmetry is spontaneously
broken and above which the chiral symmetry restores.
This result should be compared with the previous
analysis in the quenched case.
The reason why we can obtain a finite critical number
is that the $1/N_f$ expansion alters the infrared behavior
of the gauge boson propagator
from $1/k^2$ to $1/(\alpha k)$ due to the existence of
$\Pi_T^{(1)}(k)$ in the denominator.  This eliminates the
infrared divergence encountered in the previous subsection.
\par
Moreover, even in the Chern-Simons limit $\beta \rightarrow
0$, i.e., without the kinetic term for the gauge boson, the
finite critical number of flavors is obtained:
$N_f^c(\beta=0)=2$.
When $\beta\not=0$, the naive "continuum" (infinite cutoff)
limit $\Lambda/\alpha \rightarrow \infty$ leads to
$N_f^c \rightarrow 0$, i.e., no breaking of the chiral
symmetry.  However the first viewpoint explained in the
introduction that the dimensional quantity $\alpha$ gives a
natural scale of the theory implies that the relevant
physics comes essentially from the region $k \sim \alpha$
and the region
$k \gg \alpha$ does not give the essential contribution.
Hence we should normalize the physical
quantity by $\alpha$ (instead of $\Lambda_f$) and set the
upper bound of integration as $\Lambda_p \sim \alpha$.
This standpoint gives a finite and nonzero value for
$N_f^c$.

\section{Effective potential and stability}
\setcounter{equation}{0}
\subsection{chiral-symmetry violation (case (I))}
We define the dimensionless effective potential
$\tilde \Gamma[\phi, \chi]$ by
\begin{eqnarray}
 \tilde \Gamma[\phi, \chi]
 = {\pi^2 \over 2N_f} {\Gamma[\phi, \chi] \over
\Lambda^3},
\end{eqnarray}
in conformity with the definition of the dimensionless
order parameter (\ref{dimensionlessop}).
We consider the effective potential
\begin{eqnarray}
 \tilde \Gamma[\phi, \chi]
 = {1 \over 2} \tau_e \phi^2
 + {1 \over 2} \tau_o \chi^2
 + {1 \over 3} A \phi^3 + {1 \over 2} F \phi \chi^2
 + {\cal O}(\varphi^4),
 \label{potential1}
\end{eqnarray}
which satisfies the following relation associated with
the case (I):
\begin{eqnarray}
 {\delta \tilde \Gamma[\phi, \chi] \over \delta \phi}
 &\equiv& \tilde J_e
= \tau_e \phi + A \phi^2 + {1 \over 2} F \chi^2
+ {\cal O}(\varphi^3),
 \nonumber\\
{\delta \tilde \Gamma[\phi, \chi] \over \delta \chi}
&\equiv& \tilde J_o
= \tau_o \chi + F  \phi \chi  + {\cal O}(\varphi^3).
\label{sp1}
\end{eqnarray}
It is easy to see that, even when
$\tilde J_e = \tilde J_o = 0$,  a pair of eq.~(\ref{sp1})
has non-trivial solutions besides the trivial one
$\phi=\chi=0$.
First of all, we observe that a pair of solutions
such that $\chi\not=0$ and $\phi=0$ does not exist, which
is consistent with the assumption $J_e>J_o$ in case (I).

\par
There are two types of non-trivial solutions.
One type of non-trivial solutions is given by
\begin{equation}
 \chi^{(1)} = 0, \ \phi^{(1)} = - \tau_e A^{-1} > 0,
\end{equation}
where $\tau_e<0$ and $A>0$.
At this solution, the effective potential takes the value:
\begin{equation}
 \tilde \Gamma[\phi^{(1)}, \chi^{(1)}]
 = {1 \over 6} \tau_e^3 A^{-2} < 0,
\end{equation}
which is lower than the trivial one:
$\tilde \Gamma[\phi=0,\chi=0]=0$.
This solution implies the
spontaneous breaking of chiral symmetry and no spontaneous
breaking of parity.
\par
Another non-trivial solution is given by
\begin{equation}
 \phi^{(2)} = -  \tau_o F^{-1},
\ \chi^{(2)} = \sqrt{{2\tau_o(F \tau_e-A \tau_o)
\over F^3}}.
\end{equation}
The solution $\phi^{(2)}$ is positive for
$\tau_o<0$ and
$\chi^{(2)}$ is real and positive if
$\tau_o<0$ and $F \tau_e< A \tau_o $, as long as $A, F>0$.
This solution implies that the chiral symmetry and the
parity are simultaneously broken. It gives the
effective potential
\begin{equation}
 \tilde \Gamma[\phi^{(2)}, \chi^{(2)}]
 = {1 \over 6} \tau_o^2 F^{-3}
 (3F \tau_e -  2A\tau_o)
 < {1 \over 6} \tau_o^3 A F^{-3} < 0.
\end{equation}
Then this solution also gives lower effective potential than
$\tilde \Gamma[0,0]=0$.
\par
Which solution gives the absolute minimum? If the
second solution is the absolute minimum, the chiral
symmetry and the parity is spontaneously broken
simultaneously.
However we can show  that the first solution is
energetically favorable:
\begin{equation}
\tilde \Gamma[\phi^{(1)}, \chi^{(1)}]
< \tilde \Gamma[\phi^{(2)}, \chi^{(2)}] < 0,
\end{equation}
since
\begin{eqnarray}
&& \tilde \Gamma[\phi^{(1)}, \chi^{(1)}] -
\tilde \Gamma[\phi^{(2)}, \chi^{(2)}]
\nonumber\\
&=& [2|A\tau_o|^3+
|F\tau_e|^3-3|A\tau_o|^2|F\tau_e|] /[6A^2F^3] < 0,
\end{eqnarray}
in the allowed region of parameters, $A, F, \tau_e, \tau_o$.
\par
Now we consider the stability of the solution around the
respective stationary point:
$\phi=\phi^{(a)},\chi=\chi^{(a)}$ at which
${\partial \Gamma \over \partial \phi} = 0$
and ${\partial \Gamma \over \partial \chi} = 0$.
The effective potential can be expanded around the
stationary point as
\begin{eqnarray}
 \Gamma[\phi, \chi] = \Gamma[\phi^{(a)}, \chi^{(a)}]
 + {1 \over 2} (\delta \phi \ \delta \chi)
 H_\Gamma[\phi^{(a)}, \chi^{(a)}]
 \pmatrix{ \delta \phi  \cr
             \delta \chi \cr}
 +  {\cal O}((\delta\varphi)^3),
\end{eqnarray}
where $\delta \varphi$ denotes deviation
$\delta \phi$ or $\delta \chi$ from the respective
stationary point:
$\delta \phi:= \phi-\phi^{(a)}$,
$\delta \chi:= \chi-\chi^{(a)}$ and
$H_\Gamma[\phi, \chi]$ is the Hessian matrix defined by
\begin{eqnarray}
H_\Gamma[\phi, \chi]
:= \pmatrix{ {\partial^2 \Gamma \over \partial \phi^2}
&
{\partial^2 \Gamma \over \partial \phi \partial \chi}
\cr
 {\partial^2 \Gamma \over \partial \chi \partial \phi}
&
{\partial^2 \Gamma \over \partial \chi^2}
\cr} .
\end{eqnarray}
\par
For the above effective potential (\ref{potential1}),
the Hessian reads
\begin{eqnarray}
H_\Gamma[\phi, \chi]
= \pmatrix{ \tau_e+2A\phi & F\chi \cr
             F\chi & \tau_o+F\phi \cr}.
\end{eqnarray}
\par
Around the first solution,
\begin{eqnarray}
H_\Gamma[\phi^{(1)}, \chi^{(1)}]
 = \pmatrix{ -\tau_e & 0 \cr
             0 & (A\tau_o-F\tau_e)/A \cr}.
\end{eqnarray}
Note that $-\tau_e>0$ and $(A\tau_o-F\tau_e)/A>0$.
Hence $\tr H_\Gamma>0$ and the Hessian is positive definite,
i.e. $\det H_\Gamma >0$.  Then the first
solution $\phi^{(1)}$, $\chi^{(1)}$
gives the local minimum.
\par
Around the second solution,
\begin{eqnarray}
H_\Gamma[\phi^{(2)}, \chi^{(2)}]
 = \pmatrix{ (F\tau_e-2A\tau_o)/F & F\chi^{(2)} \cr
             F\chi^{(2)} & 0 \cr}.
\end{eqnarray}
The Hessian is negative definite, since
$\det H_\Gamma = - (F\chi^{(2)})^2<0$.
Hence the second solution $\phi^{(2)}$,
$\chi^{(2)}$ corresponds to the saddle point.
\par
Thus the first solution give the absolute minimum.
Therefore we can conclude that there are no stable
(absolute) minima at which both $\phi$ and $\chi$ are
simultaneously non-zero.
This conclusion agrees with the result of
Semenoff and Wijewardhana \cite{SW92}.
Moreover we have shown up to ${\cal O}(1/N_f)$ that the
chiral symmetry is broken spontaneously for
$N_f<N_f^c=K_1$,
while the spontaneous breaking of parity does
not occur.
In the broken phase the non-trivial order parameter behaves
near the critical point as
\begin{equation}
\phi^{(1)} = - \tau_e A^{-1}
 = (K_1/N_f-1) P_0^{-1} + {\cal O}(1/N_f^2),
\end{equation}
where the value $K_1$ is given by eq.~(\ref{K1}).
Note that $K_1$
is always positive and monotonically decreasing in
$\beta\Lambda/\alpha$ and has a maximum value
$2$ at the Chern-Simons limit: $\beta\Lambda/\alpha=0$.
Hence
$N_f^c$ has a finite upper bound $2$. For example, when
$\beta\Lambda/\alpha=1$,
$ K_1 = 2  \ln (2)  \cong 1.4.
$
\par

\subsection{parity violation (case (II))}
\par
Next we consider the case where $J_{1}>0, J_{2}>0$
$(J_e < J_o)$.
This case corresponds to the effective potential:
\begin{eqnarray}
 \tilde \Gamma[\phi, \chi]
 = {1 \over 2} \tau_e \phi^2
 + {1 \over 2} \tau_o \chi^2
 + {1 \over 3}( E \chi^3 + {3 \over 2}C \chi \phi^2)
+ {\cal O}(\varphi^4),
\label{potential2}
\end{eqnarray}
since this gives the following relation for the case (II):
\begin{eqnarray}
{\delta \tilde \Gamma[\phi, \chi] \over \delta \phi}
&\equiv& \tilde J_e
=  \tau_e \phi +  C  \phi \chi  + {\cal O}(\varphi^3),
 \nonumber\\
{\delta \tilde \Gamma[\phi, \chi] \over \delta \chi}
&\equiv& \tilde J_o
=  \tau_o \chi + {C \over 2} \phi^2 + E \chi^2
+ {\cal O}(\varphi^3).
\label{inv3}
\end{eqnarray}

\par
One non-trivial solution is given by
\begin{equation}
 \phi^{(1)} = 0, \ \chi^{(1)} = - \tau_o E^{-1} >0,
\end{equation}
where $\tau_o<0$ and $E>0$.
This solution gives the effective potential
\begin{equation}
 \Gamma[\phi^{(1)}, \chi^{(1)}]
 = {1 \over 6} \tau_o^3 E^{-2} < 0,
\end{equation}
which is lower than
$\Gamma[\phi=0,\chi=0]=0$.
This solution implies the
spontaneous breaking of parity and no spontaneous
breaking of chiral symmetry.
\par Another non-trivial solution is given by
\begin{equation}
 \chi^{(2)} = -  \tau_e C^{-1},
\ \phi^{(2)} = \sqrt{{2\tau_e(C \tau_o-E \tau_e)
\over C^3}}.
\end{equation}
The solution $\chi^{(2)}$ is positive for
$\tau_e<0 $ and
$\chi^{(2)}$ is real and positive if
$\tau_e<0, C \tau_o < E \tau_e$,
as long as $C, E>0$.
This solution gives the effective
potential
\begin{equation}
 \Gamma[\phi^{(2)}, \chi^{(2)}]
 = {1 \over 6} \tau_e^2 C^{-3}
 (3C \tau_o -  2 E\tau_e)
 < {1 \over 6} \tau_e^3 C^{-3} E   < 0,
\end{equation}
which is lower than
$\Gamma[0,0]=0$.
\par
We can show that the first solution is energetically
favorable:
\begin{equation}
\tilde \Gamma[\phi^{(1)}, \chi^{(1)}] <
\tilde \Gamma[\phi^{(2)}, \chi^{(2)}] < 0,
\end{equation}
in the same way as the previous case (I).
\par
For the above effective potential (\ref{potential2}),
the Hessian reads
\begin{eqnarray}
H_\Gamma[\phi, \chi]
= \pmatrix{ \tau_e+C\chi & C\phi \cr
             C\phi & 2E \cr}.
\end{eqnarray}
\par
Around the second solution,
\begin{eqnarray}
H_\Gamma[\phi^{(2)}, \chi^{(2)}]
 = \pmatrix{ 0 & C\phi^{(2)} \cr
             C\phi^{(2)} & 2E \cr}.
\end{eqnarray}
The Hessian is negative definite, since
$\det H_\Gamma = - (C\phi^{(2)})^2<0$.
Hence the second solution $\phi^{(2)}$,
$\chi^{(2)}$ corresponds to the saddle point.
\par
Around the first solution,
\begin{eqnarray}
H_\Gamma[\phi^{(1)}, \chi^{(1)}]
 = \pmatrix{ (E\tau_e-C\tau_o)/E & 0 \cr
             0 &  2E \cr}.
\end{eqnarray}
Note that $E>0, (E\tau_e-C\tau_o)/E>0$.
Hence $\tr H_\Gamma>0$ and the Hessian is positive definite,
i.e. $\det H_\Gamma >0$.
Then the first solution $\phi^{(1)}$, $\chi^{(1)}$
gives the absolute minimum.
Therefore we can conclude that there are no stable
(absolute) minima at which both $\phi$ and $\chi$ are
simultaneously non-zero.
\par
Hence  we are tempted to conclude that up to
${\cal O}(1/N_f)$ the parity symmetry may be broken
spontaneously for
$N_f<N_f^c=L_1$, while the spontaneous breaking of chiral
symmetry does not occur.
If so, the parity-odd order parameter should behave as
\begin{equation}
 \chi^{(1)} = - \tau_o E^{-1}
 = (L_1/N_f-1)S_0^{-1} + {\cal O}(1/N_f^2)>0.
\end{equation}
However note that the value
$L_1$ given by eq.~(\ref{L1}) has a maximum value 0.1208
(at $x:=\beta \Lambda/\alpha \cong 13.3$) which is extremely
small. Above which  $L_1$ decreases monotonically and goes
to zero at $x \rightarrow \infty$, while $L_1$ becomes
negative below $x=3.92$ and reaches to $L_1=-2$ at
the Chern-Simons limit: $x=0$. This seems to show that the
spontaneous breaking of parity does not occur actually.
The absence of spontaneous breakdown of parity in
(2+1)-dimensional gauge theories is consistent with results
of other works
\cite{ABKW86b,Polychronakos88,RY86,CCW91,SW92,KEIT95}
and general consideration by Vafa and Witten
\cite{VW84}.

\section{First order transition}
\setcounter{equation}{0}
\subsection{effective potential in the presence of bare
Chern-Simons  term}
In this section we study the effect of the bare Chern-Simons
term. In what follows we restrict our
consideration to small $\theta$  for simplicity of
calculation.  It does not cause any problem to consider
the large $\theta$ case, in principle besides the
calculational complexity.  Taking into account
eq.~(\ref{estimate}), we obtain from eq.~(\ref{two-op}):
\begin{eqnarray}
 \phi &:=& a_\theta J_o
 + K \tilde J_e - (P\tilde J_e^2 + Q\tilde J_o^2)
 +  {\cal O}(\theta J^2, J^3),
\nonumber\\
 \chi &:=& b_\theta + a_\theta J_e
 + L \tilde J_o - R(2\tilde J_e \tilde J_o)
 +  {\cal O}(\theta J^2, J^3),
 \label{}
\end{eqnarray}
where $a_\theta$ and $b_\theta$ are quantities of order
$\theta/N_f$ given by
\begin{eqnarray}
 a_\theta &:=&
  -  {\pi^2 \over 2\Lambda_f^2}
  {\theta \over N_f}  {32 \over \pi}
\int^{\Lambda_p}  {d^3k \over (2\pi)^3}
 {\alpha \over (\beta k^2 + \alpha k)^2}
 + {\cal O}\left({\theta^2 \over N_f^2}\right),
\nonumber\\
 b_\theta &:=&    {\pi^2 \over 2\Lambda_f^2}
 {\theta \over N_f}
 \int^{\Lambda_p}  {d^3k \over (2\pi)^3}
 {4\alpha k \over (\beta k^2 + \alpha k)^2}
 + {\cal O}\left({\theta^3 \over N_f^2}\right).
\end{eqnarray}
The inverted equation is given by
\begin{eqnarray}
 \tilde J_e &=&  D_\theta \phi + \tau_e \phi
   + A \phi^2 + B \chi^2
 +  {\cal O}(\theta \varphi^2, \varphi^3),
\nonumber\\
 \tilde J_o &=& C_\theta + E_\theta \phi + \tau_o \chi
   + F \phi \chi
 +  {\cal O}(\theta \varphi^2, \varphi^3).
 \label{invert1st}
\end{eqnarray}
{}From the consistency condition, the relations among the
coefficients are obtained in the same way as before.  For
example, we obtain:
\begin{eqnarray}
 C_\theta &=&   Q^{-1} a_\theta
 = - L^{-1}b_\theta \sim {\cal O}(\theta),
 \nonumber\\
 D_\theta &=&   a_\theta K^{-1} \tau_o
 = a_\theta K^{-1} L^{-1}(1-a_\theta^2 K^{-1} L^{-1})^{-1}
 = a_\theta K^{-1} L^{-1} + {\cal O}(\theta^3),
 \nonumber\\
 E_\theta &=&  a_\theta L^{-1} \tau_e
 = a_\theta K^{-1} L^{-1}(1-a_\theta^2 K^{-1} L^{-1})^{-1}
 = a_\theta K^{-1} L^{-1} +  {\cal O}(\theta^3),
 \nonumber\\
 \tau_e &=&
    K^{-1}  (1-a_\theta^2 K^{-1} L^{-1})^{-1}
    = K^{-1} + {\cal O}(\theta^2),
 \nonumber\\
 \tau_o &=&
  L^{-1}  (1-a_\theta^2 K^{-1} L^{-1})^{-1}
  = L^{-1} + {\cal O}(\theta^2),
\end{eqnarray}
where we have used $R=Q =-[s_1-s_2-(t_1-t_2)]$, see
eq.~(\ref{two-op}).
In the limit $\theta \rightarrow 0$, we see
$
C_\theta, D_\theta, E_\theta
\sim {\cal O}(\theta) \rightarrow 0,
$
and
$
\tau_e \rightarrow K^{-1},
\tau_e \rightarrow L^{-1},
$
as expected from the result of section 5.
As in section 7, the effective potential is obtained by
integrating  eq.~(\ref{invert1st}):
\begin{eqnarray}
 \tilde \Gamma[\phi,\chi]
= C_\theta \chi + E_\theta \phi \chi
 + {1 \over 2} \tau_e \phi^2 + {1 \over 2} \tau_o \chi^2
  + {1 \over 3} A \phi^3 + {1 \over 2} F \phi \chi^2
+ {\cal O}(\theta \varphi^3, \varphi^4),
\end{eqnarray}
so that the effective potential satisfies
\begin{eqnarray}
 {\delta \tilde \Gamma \over \delta \phi}  = \tilde J_e,
 \quad
 {\delta \tilde \Gamma \over \delta \chi}  = \tilde J_o,
\end{eqnarray}
where we have used $E_\theta=D_\theta$.
\par
Let $\phi^*,\chi^*$ denote the stationary point of the
effective potential $\tilde \Gamma[\phi,\chi]$, i.e.,
\begin{eqnarray}
 {\delta \tilde \Gamma \over \delta \phi}
 \Big|_{\phi=\phi^*,\chi=\chi^*}= 0,
 \quad
 {\delta \tilde \Gamma \over \delta \chi}
 \Big|_{\phi=\phi^*,\chi=\chi^*}= 0.
\end{eqnarray}
In order to specify the location of the stationary point, we
put
$r := \chi^*/\phi^*$.
{}From the first equation of eq.~(\ref{invert1st}), we get
\begin{eqnarray}
 \phi^*  &=&  - {\tau_e + E_\theta r \over A + B r^2}.
\end{eqnarray}
On the other hand, the second equation of
eq.~(\ref{invert1st}) leads to
\begin{eqnarray}
 \phi^*  &=&  - {E_\theta + \tau_o r \pm
 \sqrt{(E_\theta + \tau_o r)^2 - 4FC_\theta r} \over 2Fr}.
\end{eqnarray}
Equating two values for $\phi^*$ in the above two
equations, we get the algebraic equation for $r$ which is 4th
order in $r$.   Since we restrict our attention to small
$\theta$,
 $r$ is obtained up to ${\cal O}(\theta)$ as
\begin{eqnarray}
 r_\theta  &=&  A {AC_\theta-E_\theta \tau_e \over
A\tau_e\tau_o-F\tau_e}
 + {\cal O}(\theta^2),
\end{eqnarray}
since $C_\theta, E_\theta \sim {\cal O}(\theta)$.

At the stationary point we define an  order
parameter $\varphi$ by
\begin{eqnarray}
  \phi^* = {1 \over (1+r_\theta^2)^{1/2}} \varphi,
  \quad
  \chi^* = {r_\theta \over (1+r_\theta^2)^{1/2}} \varphi  .
\end{eqnarray}
In the case of $\theta=0$, $r_\theta=0$ and hence
$(\phi^*,\chi^*)=(\varphi,0)$ which is  consistent with
the result of section 7.1.
In the plane $\chi^*=r_\theta \phi^*$
including the stationary point $(\phi^*,\chi^*)$, the
effective potential behaves as (see Fig.~4)
\begin{eqnarray}
 \Gamma[\varphi]
&=& {r_\theta \over (1+r_\theta^2)^{1/2}} C_\theta
\varphi
 + {1 \over 2} {1 \over 1+r_\theta^2}
 (\tau_e + r_\theta^2 \tau_o + 2 E r_\theta )\varphi^2
 \nonumber\\
&& + {1 \over 3} {1 \over (1+r_\theta^2)^{3/2}}
 (A + {3 \over 2}r_\theta^2 F)\varphi^3
+ {\cal O}(\theta \varphi^3, \varphi^4),
\label{epcs0}
\end{eqnarray}
This is a generalization of eq.~(\ref{potential1}) to the
case of
$\theta \not= 0$.

\subsection{1st order transition}
\par

In the presence of the bare Chern-Simons term, we are
forced to investigate the behavior of the effective
potential of the form:
\begin{equation}
 \Gamma[\varphi]
 = A_1(N_f,\theta) \varphi + {1 \over 2} A_2(N_f,\theta)
\varphi^2 + {1 \over 3} A_3(N_f,\theta) \varphi^3
+ {\cal O}(\theta \varphi^3, \varphi^4) ,
\label{epcs}
\end{equation}
where all the coefficient $A_i (i=1,2,...)$ are
functions of the bare parameters: $N_f$ and $\theta$.
\par
First we consider the situation in which the Chern-Simons term
is absent: $A_1=0$ and $A_3>0$.
For positive $A_2>0$ (point A), the effective
potential is a convex function of $\varphi$   in the region
$\varphi \ge 0$ and has an absolute minimum at
$\varphi=0$, see Fig.~5
  For negative $A_2<0$ (point B), it has a minimum at
$\varphi=\varphi_m:=-A_2/A_3>0$.  This region of
parameters corresponds to the chiral symmetry-breaking
phase. As the parameter $A_2$ passes through $A_2=0$ from
the positive side to negative one, the absolute minimum at
$\varphi=0$ moves continuously to the absolute minimum at
$\varphi = \varphi_m \not= 0$. Therefore this
theory exhibits a second order phase transition at
$A_2(N_f=N_f^c,0)=0$.  This determines $N_f^c$, the
critical number of flavors, below which $(N_f<N_f^c)$ the
chiral symmetry is spontaneously broken.
\par
\par
Now we can show that the non-vanishing $A_1$ changes the
type of the phase transition drastically. In this case,
there occurs a first order phase transition at a certain
value of $A_1$.
In order to show this, we start from a point B in the
broken-symmetry phase of the phase diagram $(N_f,\theta)$
such that  $N_f<N_f^c$ and $\theta=0$.  At this point,
$A_1(N_f,\theta=0)=0$ and  $A_2(N_f,\theta=0)<0$.
As the Chern-Simons  coefficient $\theta$ grows from zero,
the shape of the effective potential $\Gamma[\varphi]$
changes as shown in Fig.~5.
For small $\theta$ (point C),  $A_1>0$ is also small and
$\Gamma[\varphi]$ has an absolute minimum at
$\varphi = \varphi_m(N_f,\theta)$ away from $\varphi=0$.
As $A_1$ is increased, the value $\Gamma[\varphi_m]$ at
$\varphi=\varphi_m$ increases monotonically and approaches
to zero at $A_1=A_1^c$ (point D). Therefore there is a
certain value of
$A_1$ such that $\Gamma[\varphi_c]=0$ at $A_1=A_1^c$ and
above which $(A_1>A_1^c)$ the absolute minimum moves to the
origin discontinuously (point E).  This is nothing but the
first order phase transition point. At the critical point,
the effective potential takes the form:
\begin{equation}
\Gamma[\varphi]
= {1 \over 3} A_3(N_f,\theta) \varphi(\varphi-\varphi_c)^2
+ {\cal O}(\varphi^4).
\label{epc}
\end{equation}
Therefore there is a critical value $\theta_c(>0)$ of
Chern-Simons  coefficient at which  the discontinuous first
order phase transition occurs in the region $N_f<N_f^c$.
By comparing eq.~(\ref{epc}) with eq.~(\ref{epcs}),  the
critical value of $\theta_c=\theta_c(N_f)$ as a function of
$N_f$ is determined:
$\theta_c$ satisfies the following equation:
\begin{equation}
 A_1(N_f,\theta)
 = {3 \over 16} {A_2^2(N_f,\theta) \over A_3(N_f,\theta)}
 ( \ll 1) ,
 \label{condition}
\end{equation}
and
$\varphi_c$ is given by
\begin{equation}
\varphi_c = - {3 \over 4} {A_2(N_f,\theta_c) \over
A_3(N_f,\theta_c)}
= \sqrt{ {3A_1(N_f,\theta_c) \over A_3(N_f,\theta_c)} } (> 0).
\end{equation}
\par
Here note that $r_\theta$ is proportional to the Chern-Simons
coefficient and hence  $\theta=0$ implies
$A_1=0$.   Therefore the first order phase transition
in Maxwell-Chern-Simons theory with fermions occurs only
when there is a bare Chern-Simons term, $\theta\not=0$.
\par
Therefore, comparing eq.~(\ref{epcs0}) with
eq.~(\ref{epcs}), we find that
\begin{eqnarray}
 A_1 &=& k_0 {\theta^2 \over N_f^2} + {\cal O}(\theta^4),
\nonumber\\
 A_2 &=&  k_1 \left(1-{N_f^c \over N_f} \right)
 + \tilde k_1 {\theta^2 \over N_f^2}
 + {\cal O}(\theta^4),
\nonumber\\
 A_3 &=& k_2 {1 \over N_f} + {\cal O}(\theta^2),
\end{eqnarray}
and hence the critical line for the first order chiral
phase transition is given by
\begin{eqnarray}
 \theta \sim \left(1-{N_f^c \over N_f} \right) ,
 \label{criticalline}
\end{eqnarray}
in the neighborhood of the critical point
$(N_f,\theta)=(N_f^c,0)$.
This result should be compared with that of Schwinger-Dyson
equation approach
\cite{KM95}:
\begin{eqnarray}
 \theta \sim \exp \left[ -
 {\pi \over \sqrt{N_f^c/N_f-1}} \right].
 \label{criticallinesd}
\end{eqnarray}
The result eq.~(\ref{criticalline}) is reasonable, since it
is obtained up to the leading order of $1/N_f$ expansion.
To the lowest order of $1/N_f$ expansion, we can not
recover the essential singularity behavior even if it is
correct, see \cite{KITE94}.

\section{Conclusion and discussion}
\setcounter{equation}{0}

In the leading order of $1/N_f$ expansion, we have
constructed the gauge-covariant
(gauge-parameter-independent) effective potential in terms
of two order parameters,
$\langle \bar \Psi \Psi \rangle$ and
$\langle \bar \Psi \tau \Psi \rangle$,
in the  Maxwell and/or Chern-Simons theory
coupled to fermionic matter. From the viewpoint of the
stability of the solution around the stationary point of the
effective potential, we have shown that the spontaneous
breaking of both the chiral-symmetry and the parity can not
occur simultaneously. This result in the absence of the
bare Chern-Simons term agrees with that obtained
only in the Landau gauge by Semenoff and Wijewardhana
\cite{SW92}.
In (2+1)-dimensional gauge theories without the bare
Chern-Simons term
$(\theta=0)$, there may exist a finite number of critical
flavors $N_f^c$ such that for
$N_f<N_f^c$ the chiral symmetry is spontaneously broken.
On the other hand, the  spontaneous breakdown of parity
does not occur. This conclusion agrees with the previous
analyses
\cite{ABKW86b,Polychronakos88,RY86,CCW91,SW92,KEIT95}
 and general consideration \cite{VW84}.
 We wish to emphasize that all the results obtained in
this paper are gauge-parameter independent.

\par
Moreover, in sharp contrast to the Schwinger-Dyson
equation approach of QED3 up to now (see e.g. \cite{KM95}),
we have included the effect coming from the vacuum
polarization to the photon propagator exactly, up to the
leading order of $1/N_f$ expansion.
In the Schwinger-Dyson  equation approach for QED3 so far,
the  vacuum polarization effect to the photon propagator has
been included through the expression
$\Pi_T^{(1)}(k)= \alpha k $ which is obtained in
the  massless fermion limit of $\Pi_T^{(1)}(k)$.
In order to include the vacuum polarization effect from
the massive fermion in the Schwinger-Dyson  equation
approach, we must solve the Schwinger-Dyson equation for the photon
propagator simultaneously with the Schwinger-Dyson equation for the
fermion propagator.
In the four-dimensional case such a calculation has been
performed in the strong-coupling phase of QED4
\cite{KMN92}.
Therefore the result obtained in this paper based on the
inversion method goes beyond the Schwinger-Dyson  equation
approach for QED3 done so far in a sense just mentioned.
However our analysis shows that the inclusion of the
massive fermion does not change the critical number of
flavors, if any.
\par
The critical value $N_f^c$ obtained here seems
rather small compared with the result of the Schwinger-Dyson
equation
$N_f^c = 3 \sim 4$ and the
Monte Carlo simulation, $N_f^c \cong 3.5 \pm 0.5$.
This will be due to the fact that the above
calculation is restricted to the leading order
of $1/N_f$ expansion, although the calculation is
systematic.
This is also the case for QED4, see \cite{KITE94}.
To estimate more accurately the critical flavor, we need
to perform the higher order calculation.

\par
 The order of the chiral phase transition is of the second
order  in the absence of the Chern-Simons term.
This is consistent with the result of \cite{SSW94}.
However, in the presence of the Chern-Simons term, we
have shown that the chiral phase transition turns into the
first order transition, as discovered firstly in the
Schwinger-Dyson equation \cite{KM95}.
\par
Now we come to the stage of discussing the meaning of the
"continuum" limit (of removing the cutoffs) in the
non-perturbative sense \cite{Creutz83,Miransky85}.
The point $(N_f^c,0)$ is the only one point which exhibits
the continuous second order chiral transition.  Therefore,
in order to obtain the continuum theory with a finite
fermion mass, the bare parameters must be adjusted so as to
approach the critical point $(N_f^c,0)$.
\par
It should be remarked that the bare Chern-Simons term plays
completely different role from the induced Chern-Simons term.  It is
the bare Chern-Simons term which causes the first order
transition, while the induced Chern-Simons term does not
leads to the first order transition, even if the
Chern-Simons term may be induced by radiative corrections
at one-loop. The Chern-Simons coefficient
$\theta$ is not subject to renormalization according to the
Coleman-Hill theorem
\cite{CH85} and the beta function of the renormalization
group is identically zero: $\beta(\theta) \equiv 0$.
The continuum theory obtained from the massive phase in this
continuum limit may have no Chern-Simons term in the
renormalized sense.
\par
In this paper we have neglected the possibility of the
fluctuation-induced first order transition.  If this
happens in this theory, the point $(N_f^c,0)$ will be no
longer of the second order transition point.  To clarify
this problem, we need to perform more careful analysis on the
infrared behavior of the theory.  However the absence of
such a first order transition has been reported recently in
\cite{SSW94}.
\par
Finally we remark the possible effect coming from higher
orders in $1/N_f$ expansion \cite{Coquereaux81}.
Our analysis of the effective potential up to the leading
order of
$1/N_f$ expansion shows
\begin{equation}
{\langle \bar \Psi \Psi \rangle \over \alpha^2}
\sim   \left( 1- {N_f^c \over N_f} \right),
\end{equation}
in QED3 without the Chern-Simons term.
In the Schwinger-Dyson  equation approach the chiral order
parameter
$\langle \bar \Psi \Psi \rangle$
as well as the dynamically generated fermion mass $m_d$
exhibits the essential singularity at the critical point
$N_f=N_f^c$:
\begin{equation}
{\langle \bar \Psi \Psi \rangle \over \alpha^2}
\sim   \left(  {m_d \over \alpha} \right)^{3/2}
  \sim  \exp \left[ - {3\pi \over \sqrt{N_f^c/N_f-1}} \right].
\end{equation}
In the four-dimensional case, we have shown \cite{KITE94}
how the result obtained in the Schwinger-Dyson  equation
approach for QED4 can be reproduced from the higher order
calculation in the inversion method.  For this, it is
helpful to enlarge the model so as to include the
four-fermion interaction.  We can expect that this scenario
also holds in the three-dimensional case, see e.g.,
\cite{CCW91}.  However, even if this conjecture is correct,
we will need the infinite order result of $1/N_f$ expansion
to recover the essential singularity exactly. A similar
feature can be seen in the expression of the critical line
between eq.~(\ref{criticalline}) and
eq.~(\ref{criticallinesd}). The result $N_f^c=2$ up to the
leading order for the Chern-Simons theory $(\beta = 0)$ with
$N_f$ flavors of 4-component fermions is not so large  that
the reliability of $1/N_f$ expansion may be not very
definite.  To see the improvement of the expansion by higher
orders, we need to calculate the next-to-leading order.
This is the subject of the subsequent work.
\par
One more comment is on the choice of order parameter.  In
the analysis of this paper, we did not consider the order
parameter
\begin{equation}
 \omega :=  {1 \over \Omega}
  \int d^3 x \langle \epsilon^{\mu\nu\rho}
 A_\mu(x) \partial_\nu A_\rho(x) \rangle .
\end{equation}
This order parameter $\omega$ as well as $\chi$ is able to
signal the spontaneous breakdown of parity after taking the
limit
$\theta \rightarrow 0$.
In this case we must obtain the effective potential as a
function of three order parameters $\phi$, $\chi$ and
$\omega$.  This is possible in principle, but is somewhat
tedious.
To discuss the connection of our result with the anyonic
superconductivity, we must include the chemical potential
into this theory.  This will be done in the forthcoming
paper.
The first order phase transition induced by the
topological term may have other important implications,
see \cite{Schierholz94}.

\section*{Acknowledgements}
This work is supported in part by the Grant-in-Aid for
Scientific Research from the Ministry of Education, Science
and Culture (No.07640377).

\appendix
\section{Clifford algebra and its representation}
\setcounter{equation}{0}

In the Minkowski space-time, the gamma matrices
$\gamma_M^\mu$ satisfy the Clifford algebra:
\begin{eqnarray}
 \gamma_M^\mu \gamma_M^\nu + \gamma_M^\nu \gamma_M^\mu
 = 2 g^{\mu \nu},
\end{eqnarray}
which is satisfied for example by
\begin{eqnarray}
 \gamma_M^0 =  \pmatrix{ \sigma_3 & 0          \cr
                       0       & -\sigma_3  \cr} ,
                       \quad
 \gamma_M^1 =  \pmatrix{ i \sigma_1 & 0            \cr
                       0         & -i \sigma_1  \cr} ,
                       \quad
 \gamma_M^2 =  \pmatrix{ i \sigma_2 & 0            \cr
                       0         & -i \sigma_2  \cr} ,
\end{eqnarray}
with Pauli matrices:
\begin{eqnarray}
 \sigma^1 =  \pmatrix{ 0  & 1  \cr
                      1  & 0  \cr} ,
                       \quad
 \sigma^2 =  \pmatrix{ 0  & -i  \cr
                      i  & 0  \cr} ,
                       \quad
 \sigma^3 =  \pmatrix{ 1  & 0  \cr
                      0  & -1  \cr} .
\end{eqnarray}
\par
 In the euclidean space, we define the gamma matrices
satisfying
\begin{eqnarray}
 \gamma_E^\mu \gamma_E^\nu + \gamma_E^\nu \gamma_E^\mu
 = -2 \delta^{\mu \nu},
\end{eqnarray}
by
$
 \gamma_E^0 := -i \gamma_M^0 ,
 \gamma_E^1 := \gamma_M^1 ,
 \gamma_E^2 := \gamma_M^2 ,
$
i.e.,
\begin{eqnarray}
 \gamma_E^0
 =  \pmatrix{ -i \sigma_3 &  0         \cr
          0           & i \sigma_3  \cr} ,
                       \quad
 \gamma_E^1
 =  \pmatrix{ i \sigma_1 &  0          \cr
            0         & -i \sigma_1    \cr} ,
                       \quad
 \gamma_E^2
 =  \pmatrix{ i \sigma_2 & 0          \cr
           0         & -i \sigma_2    \cr} .
\end{eqnarray}
Note that
\begin{eqnarray}
 (\gamma_E^\mu)^2
 = - \pmatrix{   I & 0   \cr
                 0 & I   \cr}   \quad (\mu=0,1,2)
\end{eqnarray}
and
\begin{eqnarray}
 \gamma_E^0 \gamma_E^1 \gamma_E^2   =   \pmatrix{
i\sigma_3 \sigma_1 \sigma_2 & 0      \cr
           0         & -i\sigma_3 \sigma_1 \sigma_2   \cr}
=    - \pmatrix{  I  & 0      \cr
               0  & -I   \cr} .
\end{eqnarray}
Then we define
\begin{eqnarray}
 \gamma_E^3 :=
\pmatrix{   0 & I      \cr
           I  &  0   \cr} ,
\end{eqnarray}
\begin{eqnarray}
\gamma_E^5  :=
\gamma_E^0 \gamma_E^1 \gamma_E^2  \gamma_E^3
=     \pmatrix{  0  & -I      \cr
               I  & 0   \cr} ,
\end{eqnarray}
and
\begin{eqnarray}
\tau  = \gamma_E^3  \gamma_E^5
=     \pmatrix{  I  & 0      \cr
               0  & -I   \cr} .
\end{eqnarray}
For the projection operator:
\begin{eqnarray}
 \chi_{+}  =    \pmatrix{  I  & 0      \cr
               0  & 0   \cr} ,
\quad
 \chi_{-}  =    \pmatrix{  0  & 0      \cr
               0  & I   \cr} ,
\end{eqnarray}
we obtain the following trace formulae:
\begin{eqnarray}
 \tr[\chi_{\pm}] &=&  2,
\nonumber\\
 \tr[\gamma_E^\mu \gamma_E^\nu \chi_{\pm}]
 &=&   - 2 \delta_{\mu \nu},
\nonumber\\
 \tr[\gamma_E^\mu \gamma_E^\nu \gamma_E^\rho \chi_{\pm}]
 &=&   \mp 2 \epsilon_{\mu \nu \rho},
\nonumber\\
 \tr[\gamma_E^\mu \gamma_E^\nu \gamma_E^\alpha
 \gamma_E^\beta \chi_{\pm}]
 &=&   2( \delta_{\mu \nu}\delta_{\alpha \beta}
 -  \delta_{\mu \alpha}\delta_{\nu \beta}
 + \delta_{\mu \beta}\delta_{\nu \alpha}),
\end{eqnarray}
and
\begin{eqnarray}
 \tr[\tau ] &=&  0,
\nonumber\\
 \tr[\gamma_E^\mu \gamma_E^\nu \tau]
 &=&   \mp 2 \delta_{\mu \nu},
\nonumber\\
 \tr[\tau \chi_{\pm}] &=&  \pm 2,
\nonumber\\
 \tr[\gamma_E^\mu \gamma_E^\nu \chi_{\pm}]
 &=&   - 2 \delta_{\mu \nu},
\nonumber\\
 \tr[\gamma_E^\mu \gamma_E^\nu \gamma_E^\rho \chi_{\pm}]
 &=&   \mp 2 \epsilon_{\mu \nu \rho},
\nonumber\\
 \tr[\gamma_E^\mu \gamma_E^\nu \gamma_E^\alpha
 \gamma_E^\beta \chi_{\pm}]
 &=&   2( \delta_{\mu \nu}\delta_{\alpha \beta}
 -  \delta_{\mu \alpha}\delta_{\nu \beta}
 + \delta_{\mu \beta}\delta_{\nu \alpha}) .
\end{eqnarray}

\section{Generating functional}
\setcounter{equation}{0}
For the bosonic field, $A$, the functional integration is
performed as
\begin{eqnarray}
 \int {\cal D}A \exp [-{1 \over 2} A D^{-1} A]
 = (\Det D^{-1})^{-1/2}
 = \exp \left[ {1 \over 2} \ln \Det D \right] .
\end{eqnarray}
Since the full boson propagator $D$ satisfies the
Schwinger-Dyson  equation:
\begin{eqnarray}
 D = [D^{(0)}{}^{-1}-\Pi]^{-1} =[1-D^{(0)} \Pi]^{-1} D^{(0)},
\end{eqnarray}
we have
\begin{eqnarray}
\ln \Det [D]
&=&  \Tr \ln [D]
\nonumber\\
&=& \Tr \ln [D^{(0)}] - \Tr \ln [1-D^{(0)} \Pi]
\nonumber\\
&=& \Tr \ln [D^{(0)}] +
\sum_{n=1}^\infty {1 \over n} \Tr[(D^{(0)} \Pi)^n].
\end{eqnarray}
Then the derivative takes the form:
\begin{eqnarray}
 {\partial \over \partial J}
\ln \Det [D]
&=&   {\partial \over \partial J}
\sum_{n=1}^\infty {1 \over n} \Tr[ (D^{(0)} \Pi)^{n}]
\nonumber\\
&=&   \sum_{n=0}^\infty \Tr \left[ (D^{(0)} \Pi)^{n} D^{(0)}
{\partial \over \partial J}  \Pi^{(1)} \right]
\nonumber\\
&=&   \Tr \left[ (1-D^{(0)} \Pi)^{-1} D^{(0)}
{\partial \over \partial J}  \Pi \right]
\nonumber\\
&=&  \Tr \left[ D {\partial \over \partial J}  \Pi \right].
\end{eqnarray}
Especially, for the gauge field, we obtain
\begin{eqnarray}
 {\partial \over \partial J}  \ln \Det [D]
&=&   \int {d^Dk \over (2\pi)^D}  D_{\mu\nu}^{(1)}(k)
{\partial \over \partial J}  \Pi_{\mu\nu}^{(1)}(k).
\end{eqnarray}
\par
On the other hand, for the fermionic field, $\bar \psi,
\psi$, we have
\begin{eqnarray}
 \int {\cal D}\bar \psi {\cal D}\psi
 \exp [ \bar \psi S^{-1} \psi ]
 =  \Det S^{-1}
 = \exp \left[ - \ln \Det S \right] .
\end{eqnarray}
Since the full fermion propagator $S$ satisfies the
Schwinger-Dyson equation:
\begin{eqnarray}
 S = [S^{(0){}^{-1}}-\Sigma]^{-1}
 =  [1-S^{(0)} \Sigma]^{-1} S^{(0)},
\end{eqnarray}
we obtain in the similar way to the bosonic case
\begin{eqnarray}
\ln \Det [S]
&=&  \Tr \ln [S]
\nonumber\\
&=& \Tr \ln [S^{(0)}] - \Tr \ln [1-S^{(0)} \Sigma]
\nonumber\\
&=& \Tr \ln [S^{(0)}] +
\sum_{n=1}^\infty {1 \over n} \Tr[(S^{(0)} \Sigma)^n],
\end{eqnarray}
and
\begin{eqnarray}
 {\partial \over \partial J}  \ln \Det [S]
=  {\partial \over \partial J} \Tr \ln [S^{(0)}]
+ \Tr \left[ S {\partial \over \partial J} \Sigma \right].
\end{eqnarray}
\par
In the scheme of $1/N_f$ expansion, we find
\begin{eqnarray}
  \Pi^{(1)}_{\mu\nu}
  &\sim& e^2 N_f = \alpha,
  \nonumber\\
  \Sigma^{(1)}
  &\sim& e^2 = {\alpha \over N_f}.
\end{eqnarray}
Therefore, in the leading order of $1/N_f$ expansion, we
have only to  calculate
\begin{eqnarray}
 - {\partial \over \partial J} \Tr \ln [S^{(0)}]
+ {1 \over 2}
\int {d^Dk \over (2\pi)^D}  D_{\mu\nu}^{(1)}(k)
{\partial \over \partial J}  \Pi_{\mu\nu}^{(1)}(k).
\end{eqnarray}

\section{Vacuum polarization}
\setcounter{equation}{0}
\subsection{vacuum polarization tensor}
The vacuum polarization tensor at one-loop is obtained
from
\begin{eqnarray}
\Pi_{\mu\nu}^{(1)}{}^\pm(k;-m)
&=& - e^2 \int {d^3p \over (2\pi)^3} \tr \left[
\gamma_\mu {1 \over {\not p}+{\not k}-m} \gamma_\nu {1
\over {\not p}-m} \chi_\pm \right]
\nonumber\\
&=& - e^2 \int {d^3p \over (2\pi)^3}
{\tr[\gamma_\mu({\not p}+{\not k}+m)\gamma_\nu({\not p}+m)
\chi_\pm]
\over [(p+k)^2+m^2](p^2+m^2)} .
\nonumber\\
\end{eqnarray}
Hence the vacuum polarization tensor is decomposed into two
parts:
\begin{eqnarray}
- e^2 \int {d^3p \over (2\pi)^3}
{\tr[\gamma_\mu ({\not p}+{\not k}) \gamma_\nu {\not p}
\chi_\pm]
+ m^2 \tr[\gamma_\mu\gamma_\nu \chi_\pm]
\over (p^2+m^2)[(p+k)^2+m^2]} ,
\end{eqnarray}
and
\begin{eqnarray}
&& - e^2 m \int {d^3p \over (2\pi)^3}
{ \tr[\gamma_\mu ({\not p}+{\not k}) \gamma_\nu \chi_\pm]
+  \tr[\gamma_\mu \gamma_\nu {\not p} \chi_\pm]
\over (p^2+m^2)[(p+k)^2+m^2]} .
\end{eqnarray}
The parity even part is calculated as
\begin{eqnarray}
&& - e^2 \int {d^3p \over (2\pi)^3}
{\tr[\gamma_\mu ({\not p}+{\not k}) \gamma_\nu {\not p}
\chi_\pm]
+ m^2 \tr[\gamma_\mu\gamma_\nu \chi_\pm]
\over (p^2+m^2)[(p+k)^2+m^2]}
\nonumber\\
&=& -k^2
\left( \delta_{\mu\nu}-{k_\mu k_\nu \over k^2} \right)
{e^2 \over 2\pi} \int_0^1 dx
{x(1-x) \over [m^2+x(1-x)k^2]^{1/2}}
\nonumber\\
&=& -
\left( \delta_{\mu\nu}-{k_\mu k_\nu \over k^2} \right)
{e^2 \over 2\pi} \left[ {1 \over 2} \sqrt{m^2}
+ {k^2-4m^2 \over 4k} \arctan {k \over 2\sqrt{m^2}}
\right].
\end{eqnarray}
\par
On the other hand, the odd part reads
\begin{eqnarray}
&& -e^2 m \int {d^3p \over (2\pi)^3}
{ \tr[\gamma_\mu ({\not p}+{\not k}) \gamma_\nu \chi_\pm]
+  \tr[\gamma_\mu \gamma_\nu {\not p} \chi_\pm]
\over (p^2+m^2)[(p+k)^2+m^2]}
\nonumber\\
&=&     + e^2 m
\int {d^3p \over (2\pi)^3}
{\mp 2\epsilon_{\mu\nu\rho}k_\rho
\over (p^2+m^2)[(p+k)^2+m^2]}
\nonumber\\
&=&     \mp 2e^2 m \epsilon_{\mu\nu\rho}k_\rho
\int_0^1 dx
\int {d^3p \over (2\pi)^3}
{1 \over [p^2+2xk \cdot p+xk^2+m^2]^2}
\nonumber\\
&=&     \mp {e^2 m \over 4\pi} \epsilon_{\mu\nu\rho}k_\rho
\int_0^1 {dx \over [m^2+x(1-x)k^2]^{1/2}}
\nonumber\\
&=&     \mp {e^2 m  \over 4\pi} \epsilon_{\mu\nu\rho}k_\rho
{2 \over k} \arctan {k \over 2\sqrt{m^2}},
\end{eqnarray}
where we have used
$\tr[\gamma_\mu \gamma_\nu \gamma_\rho \chi_\pm]
= \mp 2\epsilon_{\mu\nu\rho}$
and the Feynman parameter formula
${1 \over AB} = \int_0^1 dx {1 \over [xA+(1-x)B]^2}$.
\par
For the 2-component fermion, we have
\begin{eqnarray}
\Pi_T^{(1)}(k;J) &=&
- {e^2 \over 8\pi} \left[ 2|J| + {k^2-4J^2 \over k}
\arctan {k \over 2|J|} \right],
\end{eqnarray}
\begin{eqnarray}
\Pi_O^{(1)}(k;J) &=&   {e^2 \over 2\pi} J
\arctan {k \over 2|J|} ,
\end{eqnarray}
and the the derivative with respect to the source $J$ is
given by
\begin{eqnarray}
{\partial \over \partial J} \Pi_T^{(1)}(k;J)
&=& - {e^2 \over 4\pi}
\left[ sgn(J)  {8J^2 \over k^2+4J^2} - {4J \over k}
\arctan {k \over 2|J|} \right],
\end{eqnarray}
\begin{eqnarray}
{\partial \over \partial J} \Pi_O^{(1)}(k;J)
&=& {e^2 \over 4\pi} k
\left[   - {4|J| \over k^2+4J^2} + {2 \over k}
\arctan {k \over 2|J|} \right],
\end{eqnarray}
where $k:= \sqrt{k^2}$.

\subsection{power-series expansion (I)}
Using the mathematical identities:
\begin{eqnarray}
\arcsin {x \over \sqrt{1+x^2}}
=  \arctan x
= \sum_{n=1}^\infty {(-1)^{n-1} \over 2n-1} x^{2n-1}
= {\pi \over 2}- \arctan (x^{-1}),
\end{eqnarray}
we obtain the following expansions.
\par
For $|{2J \over k}| < 1$, we get
\begin{eqnarray}
\Pi_T^{(1)}(k;J)
&=& {e^2 \over 16} k \left[ -1 + 4{J^2 \over k^2}
- sgn(J) {8 \over \pi} \sum_{n=1}^\infty
(-1)^{n+1} {n \over 4n^2-1}({2J \over k})^{2n+1} \right]
\nonumber\\
&=&  {e^2 \over 16}k
\left[ - 1 + 4 {J^2 \over k^2}
- {64 \over 3\pi} {|J|^3 \over k^3}
+ {\cal O}\left( \left({J \over k}\right)^4 \right) \right],
\end{eqnarray}
\begin{eqnarray}
\Pi_O^{(1)}(k;J)
&=& {e^2 \over 4} J \left[ 1
- sgn(J) {2 \over \pi}\sum_{n=1}^\infty
(-1)^{n-1} {1 \over 2n-1}({2J \over k})^{2n-1} \right]
\nonumber\\
&=&  {e^2 \over 4} J
\left[ 1  - {4 \over \pi}{|J| \over k} + {\cal
O}\left(\left({J \over k}\right)^3 \right) \right],
\end{eqnarray}
and
\begin{eqnarray}
{\partial \over \partial J} \Pi_T^{(1)}(k;J)
&=& {e^2 \over 2} \left[ {J \over k}
- sgn(J) {2 \over \pi} \sum_{n=1}^\infty
(-1)^{n+1} {n \over 2n-1}({2J \over k})^{2n} \right]
\nonumber\\
&=&  e^2   \left[ {1 \over 2} {J \over k}
- sgn(J) {4 \over \pi} {J^2 \over k^2}
+ {\cal O}\left(\left({J \over k}\right)^4 \right) \right],
\end{eqnarray}
\begin{eqnarray}
{\partial \over \partial J} \Pi_O^{(1)}(k;J)
&=& {e^2 \over 4} \left[ 1
- sgn(J) {4 \over \pi} \sum_{n=1}^\infty
(-1)^{n-1} {n \over 2n-1}({2J \over k})^{2n-1} \right]
\nonumber\\
&=& {e^2 \over 4} \left[ 1  - {8 \over
\pi} {|J|  \over k} + {64 \over 3\pi} {|J|^3 \over k^3}  +
{\cal O}\left(\left({J \over k}\right)^4 \right) \right].
\end{eqnarray}
\subsection{power-series expansion (II)}
\par
For $|{k \over 2J}| < 1$, we obtain
\begin{eqnarray}
\Pi_T^{(1)}(k;J)
&=& sgn(J) {e^2 \over 2\pi} k \sum_{n=1}^\infty
(-1)^n {n \over 4n^2-1}({k \over 2J})^{2n-1}
\nonumber\\
&=& sgn(J) {e^2 \over 4\pi} k
\left[ - {1 \over 3}{k \over J}
+ {\cal O}\left(\left({k \over
J}\right)^3 \right) \right],
\end{eqnarray}
\begin{eqnarray}
\Pi_O^{(1)}(k;J)
&=& sgn(J) {e^2 \over 2\pi} J \sum_{n=1}^\infty
(-1)^{n-1} {1 \over 2n-1}({k \over 2J})^{2n-1}
\nonumber\\
&=&  sgn(J) {e^2 \over 4\pi} k
\left[ 1       + {\cal O}\left(\left({k \over
J}\right)^2 \right) \right],
\end{eqnarray}
and
\begin{eqnarray}
{\partial \over \partial J} \Pi_T^{(1)}(k;J)
&=& sgn(J){e^2 \over \pi}   \sum_{n=1}^\infty
(-1)^{n-1} {n \over 2n+1}({k \over 2J})^{2n}
\nonumber\\
&=& sgn(J) {e^2 \over 4\pi} \left[
 {1 \over 3} {k^2 \over J^2} + {\cal O}\left(\left({k
\over J}\right)^4 \right) \right],
\end{eqnarray}
\begin{eqnarray}
{\partial \over \partial J} \Pi_O^{(1)}(k;J)
&=&  sgn(J) {e^2 \over \pi}  \sum_{n=2}^\infty
(-1)^n {n-1 \over 2n-1}({k \over 2J})^{2n-1}
\nonumber\\
&=& sgn(J) {e^2 \over 4\pi} \left[
{1  \over 6}{k^3 \over J^3}  +
{\cal O}\left(\left({k \over J}\right)^5 \right) \right].
\end{eqnarray}

\newpage
\centerline{\large {\bf Figure Captions}}

\begin{enumerate}
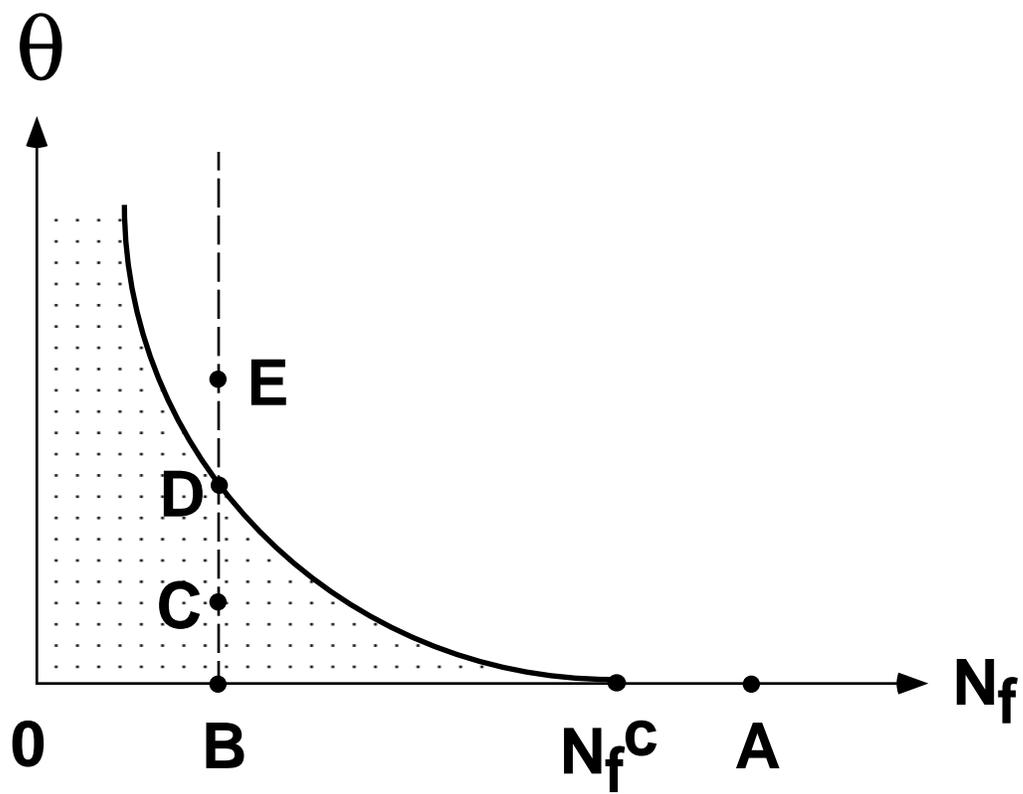
\item[Figure 1:]
Schematic phase diagram $(N_f, \theta)$.
The solid line denotes the phase transition line
below which the chiral symmetry is spontaneously broken.
\par

\begin{center}
 \begin{tabular}{|c||c|c||c|c||l|}
 \hline
 \multicolumn{6}{|c|}{Typical points in the phase diagram}
\\
   \hline   \hline
   A & $\theta=0$ & $N_f > N_f^c$ &
   $\langle \bar \Psi \Psi \rangle = 0$ &
   $\langle \bar \Psi \tau  \Psi \rangle = 0$ &
   chiral, parity symm. \\
   \hline
   B & $\theta=0$ & $N_f < N_f^c$ &
   $\langle \bar \Psi \Psi \rangle \not= 0$ &
   $\langle \bar \Psi \tau  \Psi \rangle = 0$ &
   chiral B, parity symm. \\
   \hline
   C & $0 < \theta < \theta_c$ & $N_f < N_f^c$ &
   $\langle \bar \Psi \Psi \rangle \not= 0$ &
   $\langle \bar \Psi \tau  \Psi \rangle \not= 0$ &
   chiral B, parity B \\
   \hline
   D & $\theta = \theta_c$ & $N_f < N_f^c$ &
   $\langle \bar \Psi \Psi \rangle \not= 0$ &
   $\langle \bar \Psi \tau  \Psi \rangle \not= 0$ &
   1st order transition \\
   \hline
   E & $\theta > \theta_c$ & $N_f < N_f^c$ &
   $\langle \bar \Psi \Psi \rangle = 0$ &
   $\langle \bar \Psi \tau  \Psi \rangle\not = 0$ &
   chiral symm., parity B \\
   \hline
 \end{tabular}
\end{center}

\vskip 1cm
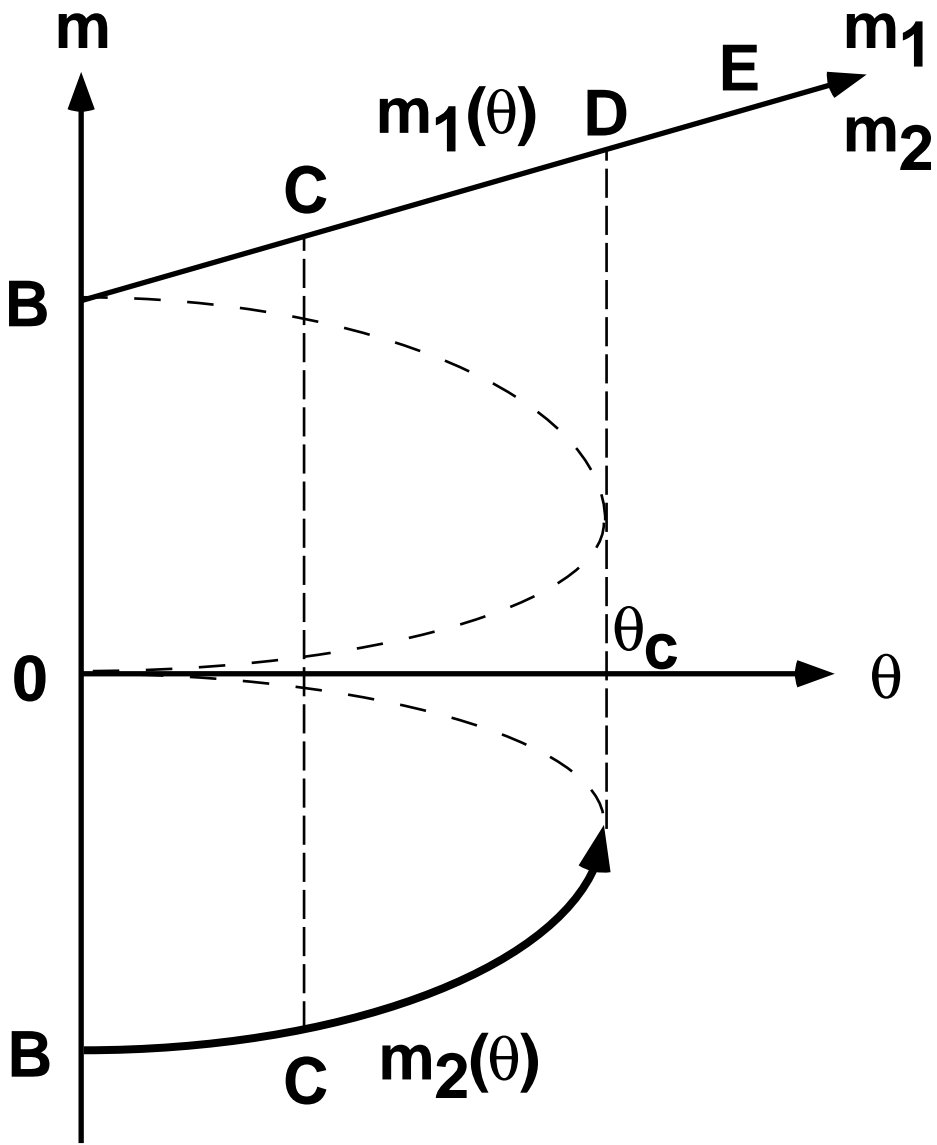
\item[Figure 2:]
Dynamical fermion masses as functions of $\theta$  for a
pair of 2-component fermions.

\vskip 1cm
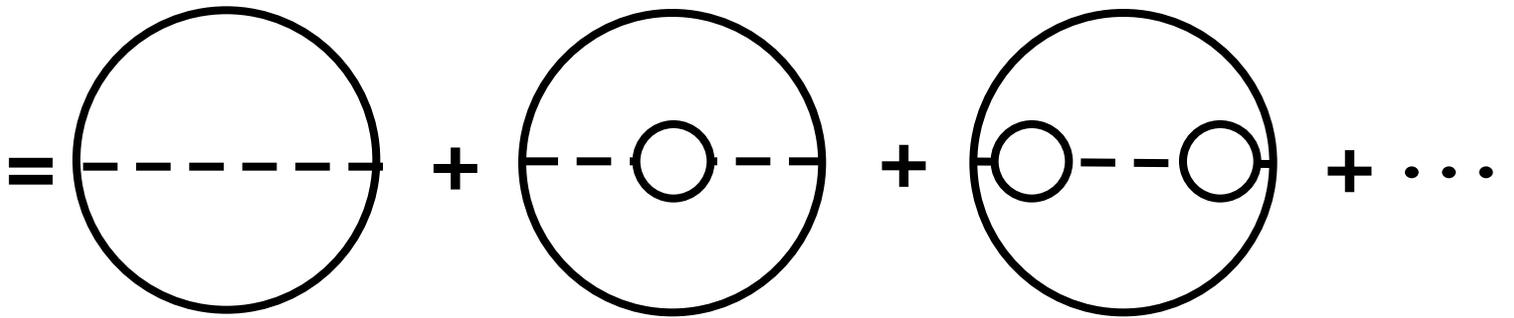
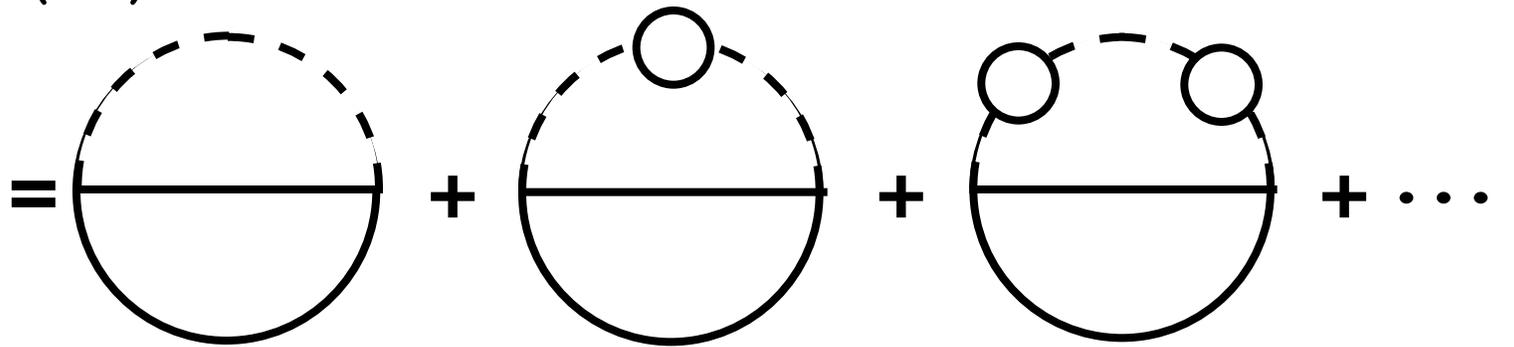
\item[Figure 3:]
Feynman diagrams needed to evaluate the effective potential
up to the leading order of $1/N_f$ expansion. The solid
line corresponds the fermion propagator and the broken line
to the gauge boson propagator.

\vskip 1cm
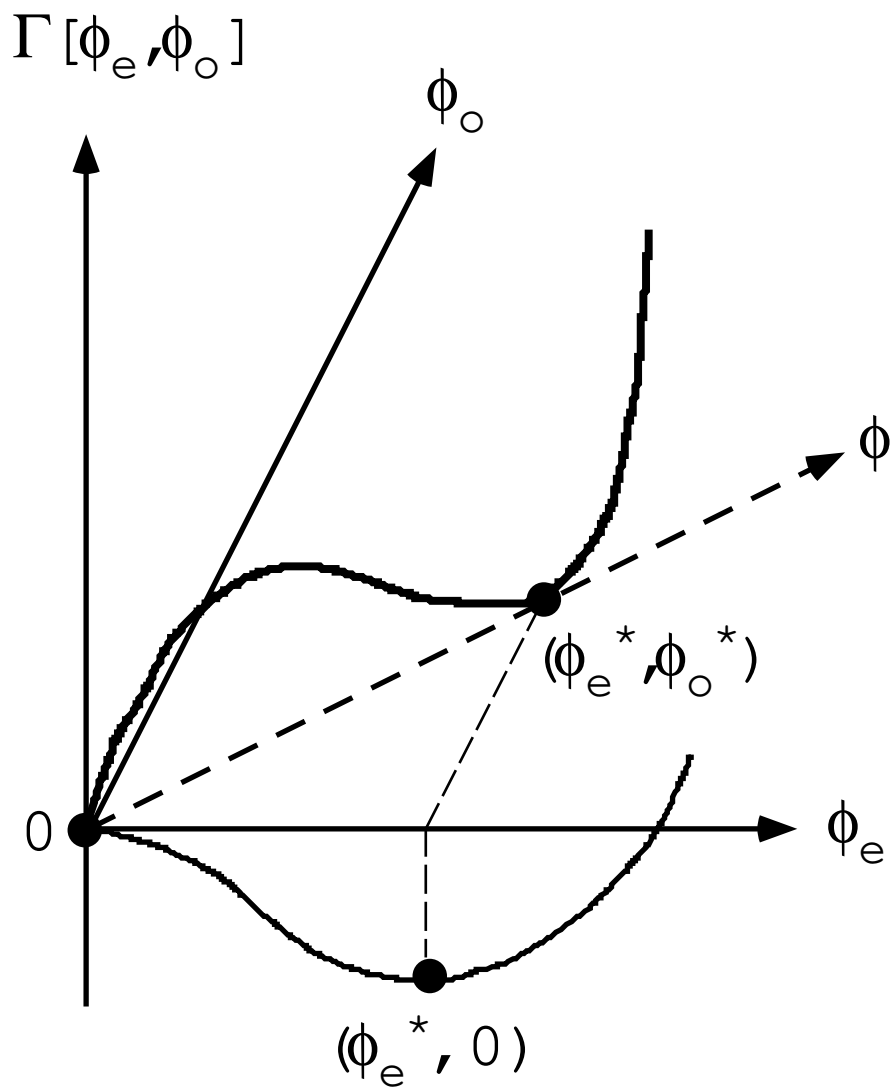
\item[Figure 4:]
The effective potential in terms of two  order parameters
$\phi=\varphi_e$ and $\chi=\varphi_o$ in the
chiral-symmetry breaking phase. When $\theta=0$ (the point
$B$), the effective potential has a minimum at
$(\varphi_e^*, 0)$.  When $\theta\not=0$, the location of
the stationary point moves to
$(\varphi_e^*,\varphi_o^*)$ where
$\varphi_e^* \not= 0$ and $\varphi_o^* \not= 0$.

\vskip 1cm
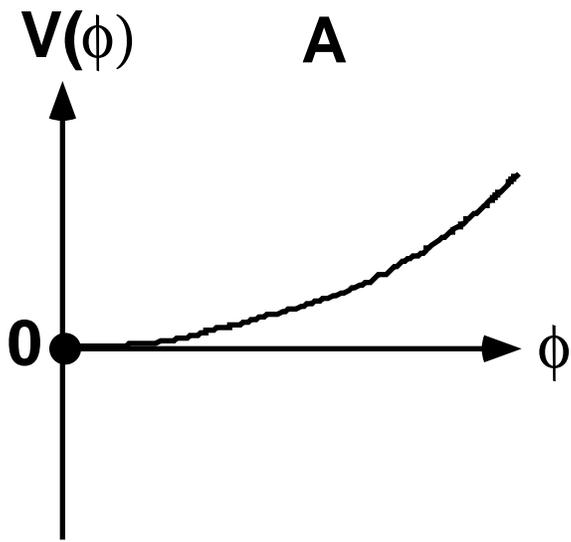
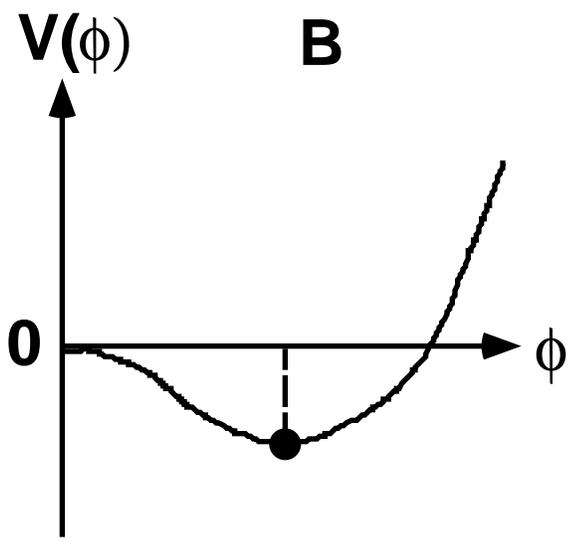
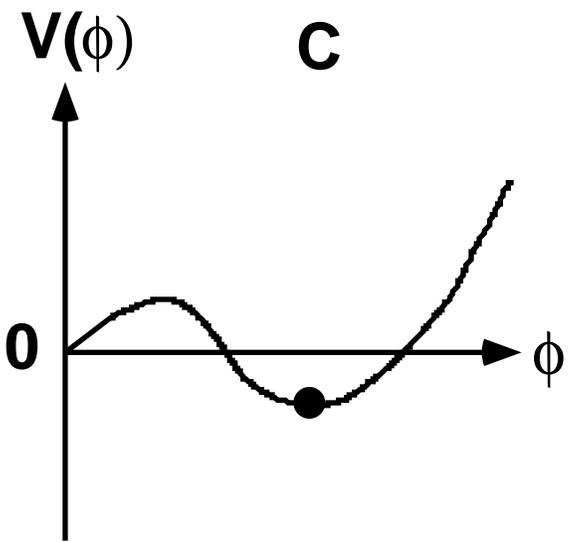
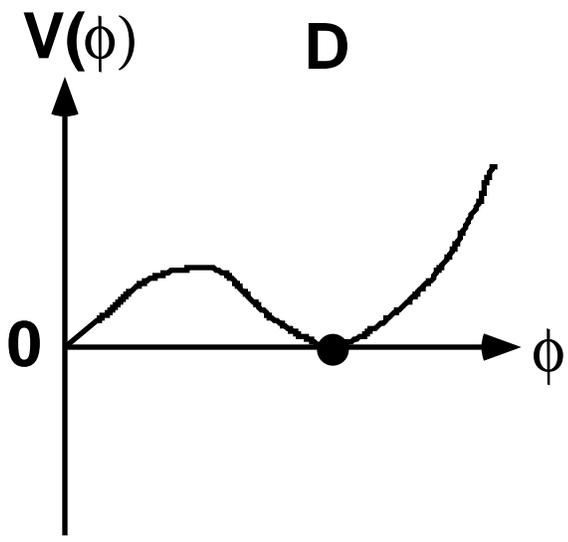
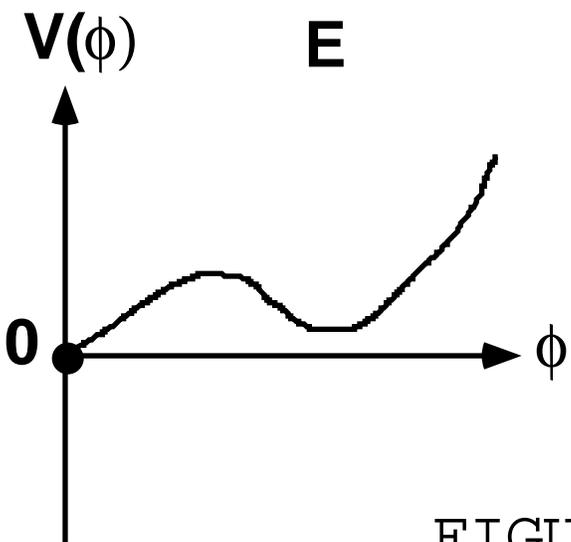
\item[Figure 5:]
The shape of the effective potential corresponding to the
respective point, $A,B,C,D,E$, in the phase diagram of
Figure 1. The phase transition from $A$ to $B$ corresponds
to the second order one, while the phase transition at $D$
is the first order one.
\end{enumerate}

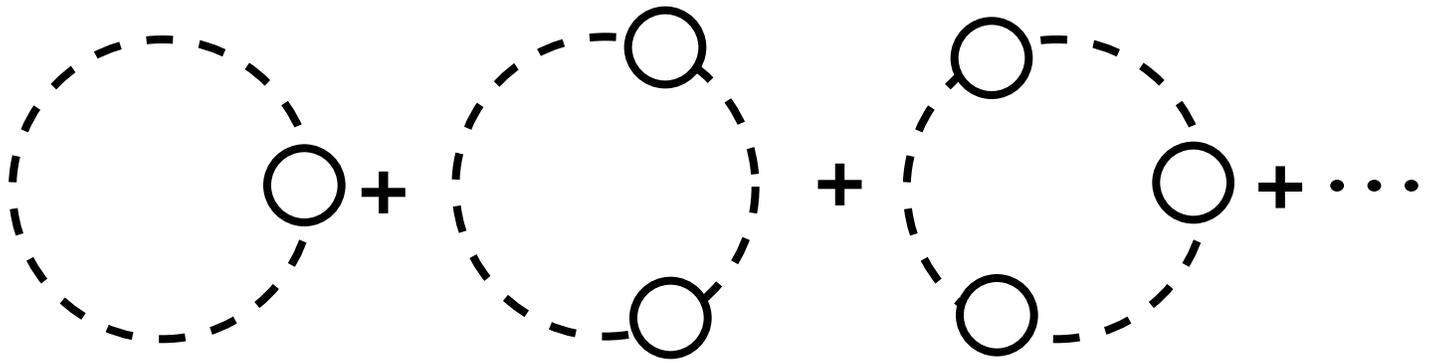
\end{document}